\documentclass[prc,notitlepage, twocolumn,
showpacs,floatfix,nofootinbib,preprintnumbers,superscriptaddress,amsmath,amssymb,longbibliography]{revtex4-2}

\usepackage[dvipdfmx]{graphicx}
\usepackage{epsfig}
\usepackage{amsmath,amssymb}
\usepackage{bm}
\usepackage{color}
\usepackage{float}
\usepackage{dcolumn} 
\usepackage{enumerate}
\usepackage{multirow}
\usepackage{threeparttable}
\usepackage[colorlinks=true,linkcolor=blue,citecolor=blue]{hyperref}
\graphicspath{{fig/}}

\begin{document}

\title{Triplet-odd pairing in finite nuclear systems: Even-even singly closed nuclei}
\author{Nobuo Hinohara}
\email{hinohara@nucl.ph.tsukuba.ac.jp}
\affiliation{
 Center for Computational Sciences, University of Tsukuba, Tsukuba, Ibaraki 305-8577, Japan
}
\affiliation{
 Faculty of Pure and Applied Sciences, University of Tsukuba, Tsukuba, Ibaraki 305-8571, Japan
}
\affiliation{Facility for Rare Isotope Beams, 
              Michigan State University, East Lansing, Michigan 48824, USA}
\author{Tomohiro Oishi}
\email{tomohiro.oishi@ribf.riken.jp}
\affiliation{RIKEN Nishina Center for Accelerator-Based Science, Wako, Saitama 351-0198, Japan}
\affiliation{Yukawa Institute for Theoretical Physics, Kyoto University, Kyoto 606-8502, Japan}
\author{Kenichi Yoshida}
\email{kyoshida@rcnp.osaka-u.ac.jp}
\affiliation{Research Center for Nuclear Physics, Osaka University, Ibaraki, Osaka 567-0047, Japan}
\affiliation{
 Center for Computational Sciences, University of Tsukuba, Tsukuba, Ibaraki 305-8577, Japan
}
\affiliation{RIKEN Nishina Center for Accelerator-Based Science, Wako, Saitama 351-0198, Japan}

\date{\today}

\begin{abstract}
\begin{description}
\item[Background] The appearance of the pairing condensate is an essential feature of many-fermion systems. There are two possible types of pairing: 
spin-singlet and spin-triplet. 
However, an open question remains as to whether the spin-triplet pairing condensate emerges in finite nuclei.
\item[Purpose] The aim of this work is to examine the coexistence 
of the spin-singlet and spin-triplet like-particle pairing condensates in nuclei. 
We also discuss the dependence on the type of pairing functional.
\item[Method] The Hartree-Fock-Bogoliubov calculations with a Skyrme $+$ local-pair energy-density functional (EDF) are performed to investigate the pairing condensate in the 
spherical ground states of Ca and Sn isotopes.
\item[Results] The spin-singlet pair EDF induces not only the spin-singlet but also the spin-triplet pairing condensates due to a strong spin-orbit splitting. 
By discarding the spin-orbit EDF, 
only the spin-singlet pairing condensate appears.
The spin-triplet pair EDF, however, induces the spin-orbit splitting and accordingly the spin-singlet pairing condensate.
\item[Conclusions] The spin-orbit splitting plays an essential role in the coexistence of the spin-singlet and spin-triplet pairing condensates in nuclei.
\end{description}
\end{abstract}

\maketitle

\section{introduction}
The pairing is universal in many-fermion systems~\cite{RevModPhys.75.607,05BB}.
A mean-field model was first introduced by Bardeen, Cooper, and Schrieffer (BCS) for describing the electronic superconductivity~\cite{57Bardeen}.
Within the original BCS theory, assumed is the condensation of a Cooper pair with
a relative angular momentum being $s$ wave,
the total spin zero,
and the center-of-mass momentum zero.
A variety of forms of pairing, unconventional pairings, that are different from the BCS type have been also found or predicted; see the reviews \cite{1991Sigrist_rev,PTPS.112.27,2003Mack_rev,Frauendorf:2014mja} and the references therein.
Especially, in electronic and cold-atomic systems,
the spin-triplet Cooper pair has been actively investigated.
The first example is the superfluidity of helium-3 atoms~\cite{1972Osheroff_01,1972Osheroff_02,1972Leggett},
where the spin-fluctuation interaction induces the spin-triplet Cooper pairs of fermionic atoms.
The Fulde-Ferrell-Larkin-Ovchinnikov type of superconductivity,
where the center-of-mass momentum is not zero,
has been also discussed for decades~\cite{1964Fulde,1964Larkin,2004Casal,2009Yanase,2023Berg}.
In several spices of heavy-fermion metals and ferromagnetic
Mott insulators \cite{2019Aoki_rev,2020Cai,doi:10.1126/science.aav8645,2020LinJiao,2022Konig},
the spin-triplet type of superconductivity is expected.
It is worthwhile to mention that spin-triplet pairing is a basic concept of topological superconductivity.
For the emergence of spin-triplet pairing, the spin-orbit interaction often plays an essential role~\cite{1991Sigrist_rev,2003Mack_rev,PhysRevLett.87.037004, PhysRevLett.107.195304,PhysRevB.84.014512}.

The pair correlation by a nucleon Cooper pair 
contributes significantly to low-energy nuclear physics. 
The BCS theory was applied to atomic nuclei soon after the original work~\cite{57Bardeen} by Bohr {\it et al.} \cite{58Bohr,69Bohr}.
The like-particle spin-singlet pairing has been investigated mostly 
and is a key to understanding the low-energy properties of the nuclear structure, for example, 
the odd-even staggering (OES) of nuclear masses, 
the collectivity of the low-lying $J^\pi=2^+$ states in even-even nuclei, 
and the moments of inertia of deformed nuclei~\cite{Ring_Schuck,13BZ}.
The unconventional pairings have also been studied in nuclear systems. 
Since a deuteron is the only two-nucleon system that is bound in nature, 
the spin-triplet proton-neutron pairing has been investigated for a long time
and is under lively discussions~\cite{Frauendorf:2014mja}. 
Similarly, due to the attractive nature of the nuclear force 
in the $^3P_2$ channel at high momentum, the emergence of 
the triplet-odd pairing has been predicted and studied 
in neutron-star matter~\cite{1968Tamagaki,1970Tamagaki,PhysRevLett.24.775}.

For the study of nuclear superfluidity in medium-mass and heavy nuclei, 
a self-consistent mean-field or energy-density functional (EDF) approach 
has been adopted~\cite{03Bender_rev}. 
This choice is advantageous as it naturally provides 
the anomalous (pair) density and the pairing gap 
as an order parameter 
and the medium effects, which are known to be strong~\cite{RevModPhys.75.607,condmat7010019}, 
can be captured in the EDF.
Since the proton-neutron spin-triplet pairing, 
as well as the isovector spin-singlet ones, can be characterized 
by local pair densities, 
there have been many studies for these types of pairing~\cite{Sagawa:2015zlu}.
On the other side, discussions on the triplet-odd pairing in nuclei
have been less active. 
In Refs.~\cite{2019OP,2021Oishi_M1},
the connection between the magnetic-dipole excitation to the triplet-odd pairing is suggested.
However, experimental evidence of the triplet-odd pairing has not been observed.

In this work, we study the like-particle spin-triplet superfluidity 
in an EDF approach.
To this end, we introduce the spin-triplet nonlocal pair density as an 
order parameter. 
We also investigate the connection between spin-orbit splitting and triplet-odd pairing.

\section{formalism}
We describe the spin-singlet and spin-triplet pairing within the local 
density approximation of the EDF.
Details on this framework are well summarized in Refs. \cite{72Vautherin,03Bender_rev,2004Jacek},
and in particular, we focus on the pairing part in this section.

\subsection{Nonlocal pair density}

The building block of the pairing in the Hartree-Fock-Bogoliubov (HFB) theory is the 
pair density matrix. We define it with the standard phase 
for the case without the proton-neutron pairing as
\begin{align}
 \hat{\tilde{\rho}}(\bm{r}_1s_1,\bm{r}_2s_2;t) 
    &= -2s_2\langle \Psi|
    \hat{c}_{\bm{r}_2-s_2t}
    \hat{c}_{\bm{r}_1s_1t}
    |\Psi\rangle,
\end{align}
where $\hat{c}_{\bm{r}st}$ represents the nucleon annihilation operator 
at position $\bm{r}$, spin $s$, and isospin $t$, 
and $|\Psi\rangle$ is the HFB state.

The spin-singlet and spin-triplet nonlocal pair densities are defined by
\begin{align}
    \tilde{\rho}_t(\bm{r}_1,\bm{r}_2) &= \sum_s
    \hat{\tilde{\rho}}(\bm{r}_1s,\bm{r}_2s;t), \label{eq:non-local-rho}\\
    \tilde{\bm{s}}_t(\bm{r}_1,\bm{r}_2) &= \sum_{s_1s_2}
    \hat{\tilde{\rho}}(\bm{r}_1s_1,\bm{r}_2s_2;t) \hat{\bm{\sigma}}_{s_2s_1}.
    \label{eq:non-local-s}
\end{align}
One can express the pair density matrix as 
\begin{align}
 \hat{\tilde{\rho}}(\bm{r}_1s_1,\bm{r}_2s_2;t)
 = \frac{1}{2}\tilde{\rho}_t(\bm{r}_1,\bm{r}_2) 
 \delta_{s_1s_2} 
 + \frac{1}{2} \tilde{\bm{s}}_t(\bm{r}_1,\bm{r}_2)
 \cdot \hat{\bm{\sigma}}_{s_1s_2},
\end{align}
where $\hat{\bm{\sigma}}$ is the spin Pauli matrix.
We note that the nonlocal pair densities show the spatial property of the 
nucleon pair; the spin-singlet pair is symmetric
and the spin-triplet pair is antisymmetric for the exchange 
of the coordinate variables,
\begin{align}
\tilde{\rho}_t(\bm{r}_1,\bm{r}_2) &= 
\tilde{\rho}_t(\bm{r}_2,\bm{r}_1), \\
\tilde{\bm{s}}_t(\bm{r}_1,\bm{r}_2) &= 
-\tilde{\bm{s}}_t(\bm{r}_2,\bm{r}_1).
\end{align}
The spin-triplet nonlocal pair density vanishes at $\bm{r}_1=\bm{r}_2$,
indicating that a simple local density approximation does not work for the spin-triplet pair, and the nonlocality plays a major role here.

\subsection{Density matrix expansion}

The nuclear interaction energy derived from a local two-body interaction 
can be expressed with the local densities 
by a density matrix expansion technique.
This has been discussed in Ref.~\cite{03Bender_rev}
for the particle-hole part of the interaction.
Here we apply the density matrix expansion for the 
pair density matrix (nonlocal pair density).

We introduce the following coordinates of the pair:
\begin{align}
    \bm{r}_1 = \bm{r} + \frac{\bm{r}_{\rm rel}}{2}, \quad
    \bm{r}_2 = \bm{r} - \frac{\bm{r}_{\rm rel}}{2}, 
\end{align}
assuming that the pair density matrix vanishes quickly with increasing 
$\bm{r}_{\rm rel}$. This allows us to expand the nonlocal pair densities 
in terms of the relative coordinate $\bm{r}_{\rm rel}$:
\begin{align}
    \tilde{\rho}_t(\bm{r}_1,\bm{r}_2) 
    &= \tilde{\rho}_t
    \left(\bm{r} + \frac{\bm{r}_{\rm rel}}{2},
    \bm{r}-\frac{\bm{r}_{\rm rel}}{2}\right) \nonumber \\
    &= \tilde{\rho}_t(\bm{r},\bm{r}) \nonumber \\
    &\quad + \frac{\partial}{\partial \bm{r}_{\rm rel}}
    \tilde{\rho}_t  \left(\bm{r}+ \frac{\bm{r}_{\rm rel}}{2},
    \bm{r}-\frac{\bm{r}_{\rm rel}}{2}\right)\Big|_{\bm{r}_{\rm rel}=\bm{0}}
    \cdot \bm{r}_{\rm rel}  \nonumber \\ &\quad 
    + \frac{1}{2}
    \frac{\partial^2}{\partial \bm{r}_{\rm rel}^2}
    \tilde{\rho}_t  \left(\bm{r} + \frac{\bm{r}_{\rm rel}}{2},
    \bm{r}-\frac{\bm{r}_{\rm rel}}{2}\right)\Big|_{\bm{r}_{\rm rel}=\bm{0}}
    \bm{r}_{\rm rel}^2 \nonumber \\ &\quad 
    +{\cal O}(|\bm{r}_{\rm rel}|^3)\nonumber \\
    &= \tilde{\rho}_t(\bm{r})
    + \frac{1}{2}( \bm{\nabla}_1 - \bm{\nabla}_2)
    \tilde{\rho}_t(\bm{r}_1,\bm{r}_2)\Big|_{ \bm{r}_1=\bm{r}_2=\bm{r}} \cdot \bm{r}_{\rm rel} \nonumber \\ &\quad 
    + \frac{1}{8}
    (\bm{\nabla}_1 - \bm{\nabla}_2)^2 
    \tilde{\rho}_t(\bm{r}_1,\bm{r}_2)\Big|_{ \bm{r}_1=\bm{r}_2=\bm{r}}  \bm{r}_{\rm rel}^2 \nonumber \\
    &\quad
    +{\cal O}(|\bm{r}_{\rm rel}|^3)
    \nonumber \\
    &= \tilde{\rho}_t(\bm{r}) 
    + \frac{1}{8} \left[ \Delta\tilde{\rho}_t(\bm{r})
    - 4 \tilde{\tau}_t(\bm{r}) 
    \right]
    \bm{r}_{\rm rel}^2  + {\cal O}(|\bm{r}_{\rm rel}|^3).
\end{align}
We use $\tilde{\rho}_t(\bm{r}_1,\bm{r}_2) = 
\tilde{\rho}_t(\bm{r}_2,\bm{r}_1)$ to remove the first-order term, and the local pair density and kinetic pair density are defined as
\begin{align}
    \tilde{\rho}_t(\bm{r}) &= \tilde{\rho}_t(\bm{r},\bm{r}), \\
    \tilde{\tau}_t(\bm{r}) &= 
    \left(\bm{\nabla}_1\cdot\bm{\nabla}_2\right)
    \tilde{\rho}_t(\bm{r}_1,\bm{r}_2)\Big|_{\bm{r}_1=\bm{r}_2=\bm{r}}.
\end{align}
The spin-triplet nonlocal pair density is expanded as
\begin{align}
\tilde{\bm{s}}_t(\bm{r}_1,\bm{r}_2) 
&= \tilde{\bm{s}}_t
\left(\bm{r}+\frac{\bm{r}_{\rm rel}}{2},\bm{r}- \frac{\bm{r}_{\rm rel}}{2}\right) 
\nonumber \\
&= 
\bm{r}_{\rm rel}\cdot
\left[
\frac{\partial}{\partial \bm{r}_{\rm rel}}\otimes
\tilde{\bm{s}}_t\left(\bm{r}+\frac{\bm{r}_{\rm rel}}{2},\bm{r}- \frac{\bm{r}_{\rm rel}}{2}\right) 
\right]_{\bm{r}_{\rm rel}=\bm{0}} \nonumber\\&\quad  
+ {\cal O}(|\bm{r}_{\rm rel}|^2) \nonumber \\
&= \frac{1}{2} \bm{r}_{\rm rel}\cdot
\left[(\bm{\nabla}_1 - \bm{\nabla}_2) \otimes
\tilde{\bm{s}}_t(\bm{r}_1,\bm{r}_2) \right]\Big|_{\bm{r}_1=\bm{r}_2=\bm{r}} \nonumber \\
&\quad + {\cal O}(|\bm{r}_{\rm rel}|^2) \nonumber \\
&= i \bm{r}_{\rm rel}\cdot \tilde{\sf J}_t(\bm{r})
+ {\cal O}(|\bm{r}_{\rm rel}|^2),
\end{align}
where 
\begin{align}
\tilde{\sf J}_t(\bm{r}) = \frac{1}{2i}
(\bm{\nabla}_1-\bm{\nabla}_2)\otimes
\tilde{\bm{s}}_t(\bm{r}_1,\bm{r}_2)
\Big|_{\bm{r}_1=\bm{r}_2=\bm{r}}
\end{align}
is the tensor (spin-current) pair density,
we use $\tilde{\bm{s}}_t(\bm{r}_1,\bm{r}_2)
= - \tilde{\bm{s}}_t(\bm{r}_2,\bm{r}_1)$
to remove the zeroth-order term,
and $\bm{v}\cdot(\bm{u}\otimes\bm{w})
\equiv (\bm{v}\cdot\bm{u}) \bm{w}$.
Spin-triplet anisotropic pairing in condensed-matter physics
requires an odd wave-number $\bm{k}$ dependence, 
and the tensor pair density corresponds to the order parameters of the $p$-wave superfluidity that consists of nine components \cite{1991Sigrist_rev}.

The density matrix expansion of the nonlocal pair density
provides the local EDF starting from a local 
two-body spin-singlet and spin-triplet pairing interaction.
The general form of the pair EDF is given by
\cite{Bennaceur_2017}
\begin{align}
E_{{\rm pair},t}^{S=0}
= \int d\bm{r}_1 d\bm{r}_2
v^{S=0}_{{\rm pair},t}(|\bm{r}_1-\bm{r}_2|)
|\tilde{\rho}_t(\bm{r}_1,\bm{r}_2)|^2, \\
E_{{\rm pair},t}^{S=1}
= \int d\bm{r}_1 d\bm{r}_2
v^{S=1}_{{\rm pair},t}(|\bm{r}_1-\bm{r}_2|)
|\tilde{\bm{s}}_t(\bm{r}_1,\bm{r}_2)|^2, \label{eq:nonlocal_spintriplet_pair}
\end{align}
where $v^{S=0}_{{\rm pair},t}$ and 
$v^{S=1}_{{\rm pair},t}$ 
are the spin-singlet and spin-triplet pairing interaction 
strengths that depend on the absolute value of the relative coordinate of the two nucleons.

Inserting the density matrix expansion in the nonlocal 
pair densities, we have 
\begin{align}
E^{S=0}_{{\rm pair},t} &= \int d\bm{r}
\left\{ \tilde{C}^\rho_t |\tilde{\rho}_t(\bm{r})|^2
+ \tilde{C}^{\Delta\rho}_t {\rm Re}\left[\tilde{\rho}^\ast_t(\bm{r}) \Delta\tilde{\rho}_t(\bm{r})\right] \right. 
\label{eq:spin-singletpairEDF}
\nonumber \\
& \quad \left. + \tilde{C}^\tau_t {\rm Re}\left[\tilde{\rho}^\ast_t(\bm{r})
\tilde{\tau}_t(\bm{r})\right]\right\},\\
E^{S=1}_{{\rm pair},t}  
&= \int d\bm{r} \tilde{C}^J_t |\tilde{\sf J}_t(\bm{r})|^2.
\end{align}
The coupling constants are related to the 
local potential as
\begin{align}
\tilde{C}^\rho_t &= \int d\bm{r}_{\rm rel} v^{S=0}_{{\rm pair},t}(|\bm{r}_{\rm rel}|), \\
\tilde{C}^{\Delta\rho}_t &= -\frac{1}{4} \tilde{C}^\tau_t = \frac{1}{4}\int d\bm{r}_{\rm rel}
\bm{r}_{\rm rel}^2 v^{S=0}_{{\rm pair},t}(|\bm{r}_{\rm rel}|), 
\label{eq:CtDrho}
\\
\tilde{C}_t^J &= \int d\bm{r}_{\rm rel} \bm{r}_{\rm rel}^2
v^{S=1}_{{\rm pair},t}(|\bm{r}_{\rm rel}|).
\end{align}
These are the coupling constants for the 
spin-singlet pairing $\tilde{C}^\rho_t$ and its next-order terms $\tilde{C}^{\Delta\rho}_t$ and $\tilde{C}^\tau_t$, 
and the spin-triplet coupling constant $\tilde{C}^J_t$.

By using the G3RS-$^1$E-1 potential introduced by Tamagaki \cite{1968Tamagaki}, for instance,
$\tilde{C}^\rho_t=-697.087$ MeV fm$^3$ and
$\tilde{C}^{\Delta\rho}_t=1363.253$ MeV fm$^5$ are obtained for the spin-singlet coupling constants.
For the spin-triplet coupling, on the other hand,
by using the G3RS-$^3$O-1 potential for $v^{S=1}_{{\rm pair},t}$, we obtain
$\tilde{C}^J_t =6794.724$ MeV fm$^5$.
Note that we assumed the $^3 P_{1}$ channel to obtain this value and that this is repulsive in this channel.

In the lowest order in terms of the nonlocality,
the local pair density $\tilde{\rho}_t(\bm{r})$ and the pair EDF 
that is proportional to $|\tilde{\rho}_t(\bm{r})|^2$ represent
the spin-singlet pair condensation and EDF, 
while the spin-current pair density $\tilde{\sf J}_t(\bm{r})$ and 
the term proportional to $|\tilde{\sf J}_t(\bm{r})|^2$ represent
the spin-triplet pair condensation and EDF.

Zero-range Skyrme interactions also produce the terms related to the spin-singlet and spin-triplet pair condensation.
Only the SkP interaction \cite{1984Jacek} includes the spin-singlet and spin-triplet terms,
and other standard Skyrme EDFs do not consider pairing terms other than those proportional to $|\tilde{\rho}_t(\bm{r})|^2$ due to 
unrealistic pairing properties \cite{03Bender_rev}.

We note that the spin-triplet pair density and thus the EDF can be decomposed into trace (pseudoscalar), antisymmetric (vector), and symmetric (pseudotensor) parts \cite{2004Jacek}:
\begin{align}
\tilde{J}_t(\bm{r}) &= \sum_{a=x,y,z} \tilde{\sf J}_{taa}(\bm{r}), \\
\tilde{\bm{J}}_{ta}(\bm{r}) &= \sum_{b,c=x,y,z} \epsilon_{abc}\tilde{\sf J}_{tbc}(\bm{r}), \\
\tilde{\underline{\sf J}}_{tab}(\bm{r}) &= \frac{1}{2}\tilde{\sf J}_{tab}(\bm{r}) + \frac{1}{2}\tilde{\sf J}_{tba}(\bm{r}) - \frac{1}{3}\tilde{J}_{t}(\bm{r}) \delta_{ab}.
\end{align}
This decomposition of the spin-current quantity is also applied in the discussion of $^3P_2$ superfluidity \cite{PhysRevD.5.1883,PhysRevResearch.2.013193},
and these three components are relevant to $^3P_0$, $^3P_1$, and $^3P_2$
superfluidity, respectively.

For a general pair EDF that is not based on an effective interaction, 
each coupling constant in the spin-singlet pair EDF (\ref{eq:spin-singletpairEDF}) can be taken independently,
except that the relation between $\tilde{C}_t^{\Delta\rho}$ and $\tilde{C}_t^\tau$ in Eq.~(\ref{eq:CtDrho}) is a requirement from the local gauge invariance \cite{2004Jacek}.
The spin-triplet pair EDF has a structure similar to that of the tensor functional in the particle-hole EDF and can have a more general form \cite{PhysRevC.80.064302}:
\begin{align}
 E^{S=1}_{{\rm pait},t} = \int d\bm{r} \tilde{C}_t^{J0}|\tilde{J}_t(\bm{r})|^2
 + \tilde{C}_t^{J1}|\tilde{\bm{J}}_t(\bm{r})|^2
 + \tilde{C}_t^{J2}|\tilde{\underline{\sf J}}_t(\bm{r})|^2.
\end{align}
Unlike the particle-hole part, these three coupling constants
are not constrained by the local gauge invariance and can be taken independently.
When the pair EDF is derived from a nonlocal effective interaction
of the form (\ref{eq:nonlocal_spintriplet_pair}),
three coupling constants are related by $\tilde{C}^J_t= 3\tilde{C}_t^{J0}
= 2\tilde{C}_t^{J1} = \tilde{C}_t^{J2}$.
However, the tensor and spin-orbit interactions, which are not in the form of 
Eq.~(\ref{eq:nonlocal_spintriplet_pair}),
allow independent contributions to the three coupling constants.

\subsection{Mean-field approach}
The mean-field equations for protons and neutrons,
obtained by the functional derivative of the 
EDF, are given in Refs. \cite{1984Jacek,2004Jacek}.
The pair Hamiltonian has the following form:
\begin{align}
\tilde{h}^{(t)}_{ss'}(\bm{r}) &= 
\left[\tilde{U}_t(\bm{r})
- \bm{\nabla} \tilde{M}_t(\bm{r})\cdot\bm{\nabla}\right]\delta_{ss'} \nonumber \\
&\quad + \frac{1}{2i}\left\{ \bm{\nabla}\cdot\left[ \tilde{\sf B}_t(\bm{r})\cdot\hat{\bm{\sigma}}_{ss'} \right]
+ \left[ \tilde{\sf B}_t(\bm{r})\cdot\hat{\bm{\sigma}}_{ss'}\right]\cdot\bm{\nabla}\right\},
\label{eq:pairHamiltonian}
\end{align}
where the potential energy $\tilde{U}_t$, the effective inertia parameter $\tilde{M}_t$, and the spin-orbit form factors $\tilde{\sf B}_t$ are given by 
\begin{align}
\tilde{U}_t(\bm{r}) &= 2\tilde{C}_t^\rho \tilde{\rho}_t(\bm{r}) + 2\tilde{C}^{\Delta\rho}_t\Delta\tilde{\rho}_t(\bm{r})
+ \tilde{C}_t^\tau \tilde{\tau}_t(\bm{r}), \\
    \tilde{M}_t(\bm{r}) &= \tilde{C}_t^\tau \tilde{\rho}_t(\bm{r}),\\ 
    \tilde{\sf B}_{tab}(\bm{r}) &= 2\tilde{C}_t^{J0}\tilde{J}_t(\bm{r})\delta_{ab}
    - 2\tilde{C}_t^{J1}\sum_{c=x,y,z} \epsilon_{acb}\tilde{\bm{J}}_{tc}(\bm{r})
    \nonumber \\ &\quad
    + 2\tilde{C}_t^{J2} \tilde{\underline{\sf J}}_{tab}(\bm{r}).
\end{align}
As the pair Hamiltonian (\ref{eq:pairHamiltonian}) depends on the spins $s$ and $s'$ when spin-triplet pair EDF is considered, we define the pairing gap by averaging out the pair Hamiltonian using the spin-dependent lower component of the quasiparticle wave function $\phi_2^{(t)}(\mu,\bm{r}s)$
as
\begin{align}
    \Delta_t &= \frac
    { \displaystyle
        \int d\bm{r} \sum_{ss'\mu} \phi^{(t)\ast}_{2}(\mu,\bm{r}s') \tilde{h}_{s's}^{(t)}(\bm{r})\phi^{(t)}_{2}(\mu,\bm{r}s)
    }
    { \displaystyle
       \int d\bm{r} \sum_{s\mu} |\phi^{(t)}_{2}(\mu,\bm{r}s)|^2 
    } \nonumber \\
    &= \frac{1}{N_t} 
    \int d\bm{r} \left[ \tilde{U}_t(\bm{r}){\rho}_t(\bm{r})
    +\tilde{M}_t(\bm{r}){\tau}_t(\bm{r})
    +\tilde{\sf B}_t(\bm{r})\cdot {\sf J}_t(\bm{r}) \right],
    \label{eq:gap}
\end{align}
where $N_t=N$ or $Z$, and
the density $\rho_t$, the kinetic density $\tau_t$, and the tensor (spin-current) density ${\sf J}_t$ are defined using the nonlocal particle-hole densities [defined in a similar way as Eqs.~ (\ref{eq:non-local-rho}) and (\ref{eq:non-local-s}) but for the particle-hole density matrix] as
\begin{align}
    \rho_t(\bm{r}) &= \rho_t(\bm{r},\bm{r}), \\
    \tau_t(\bm{r}) &= (\bm{\nabla}_1\cdot\bm{\nabla}_2) \rho_t(\bm{r}_1,\bm{r}_2)\Big|_{\bm{r}_1=\bm{r}_2=\bm{r}},\\
    {\sf J}_t(\bm{r}) &= \frac{1}{2i} (\bm{\nabla}_1-  \bm{\nabla}_2)\otimes \bm{s}_t(\bm{r}_1,\bm{r}_2)\Big|_{\bm{r}_1=\bm{r}_2=\bm{r}}.
\end{align}
Although Eq.~(\ref{eq:gap}) is a natural extension of the average gap 
for a generalized pair Hamiltonian
\cite{Bender2000},
the discrepancy of the pairing gap and experimental OES has been pointed out when the singlet-pair EDF contains the kinetic terms $\tilde{C}^{\tau}_t$ and $\tilde{C}^{\Delta\rho}_t$ \cite{0954-3899-45-2-024004}. 
\subsection{Expression within spherical symmetry}

We assume the spherical symmetry for the HFB state for simplicity.
The spherical symmetry cancels the two of the spin-current pair densities $\tilde{J}_t(\bm{r})$ and $\tilde{\underline{\sf J}}_t(\bm{r})$,
and only the radial component of the vector spin-current pair density can exist,
$\tilde{\bm{J}}_t(\bm{r}) = \tilde{J}_{tr}(r) \bm{e}_r$ \cite{PhysRevC.81.014313}.
Within the spherical symmetry, the quasiparticle wave function can be 
decomposed into the radial and angular part 
\begin{align}
 \phi_i^{(t)}(E,\bm{r}s) = \frac{ u_i^{(t)}(nlj,r)}{r} Y_{l m_l}(\bm{\hat{r}}) \langle l m_l \frac{1}{2} s | jm\rangle
 \quad (i = 1, 2),
 \label{eq:qpwave}
\end{align}
where $i=1$ and 2 correspond to the upper and lower components, respectively.
The local pair density and the spin-current pair density are given by
\begin{align}
\tilde{\rho}_t(r) &= -\frac{1}{4\pi r^2} \sum_{nlj}
(2j+1) u_1^{(t)}(nlj,r) u_2^{(t)}(nlj,r), \label{eq:rhot_sph}\\
\tilde{J}_{tr}(r) &=
-\frac{2}{4\pi r^3} \sum_{nlj}
(2j+1) \langle \bm{l}\cdot \bm{s}\rangle
 u_1^{(t)}(nlj,r) u_2^{(t)}(nlj,r), \label{eq:Jt_sph}
\end{align}
where $\langle \bm{l}\cdot\bm{s}\rangle =
\displaystyle\frac{1}{2}\left[
j(j+1)-l(l+1) - \displaystyle\frac{3}{4}\right]$.
Notice that these quantities have different dimensions.

\begin{figure}[H]
\begin{center}
\includegraphics[width=80mm]{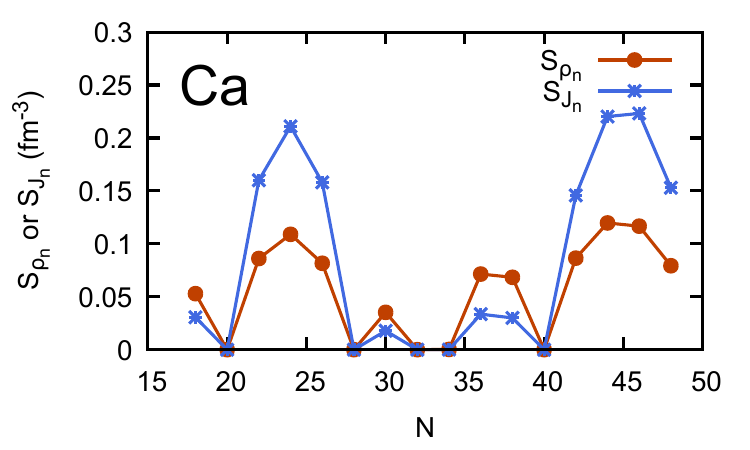} \\
\includegraphics[width=80mm]{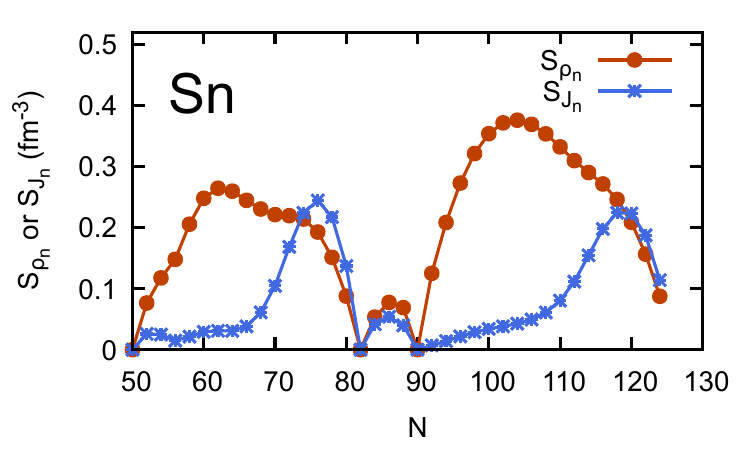} 
\end{center}
\caption{Spin-singlet and triplet pairing components of neutrons, $S_{\rho_n}$ and $S_{J_n}$, respectively, in the Ca and Sn isotope chains.}\label{fig:alice}
\end{figure}

\section{Numerical calculations}

\subsection{Spin-singlet pair EDF}

We utilize the {\sc hfbrad} code \cite{HFBRAD} for spherical Skyrme-HFB calculations in the following.
The SLy4 and spin-singlet volume-type contact pair EDF with the strength 
$\tilde{C}_n^{\rho} = -46.625$ MeV fm$^3$
(in the standard notation $V_0=4\tilde{C}_n^\rho=-186.5$ MeV fm$^3$)
is employed with the cutoff parameter of 60 MeV.
This strength has been adjusted to reproduce the neutron pairing gap 1.245 MeV in $^{120}$Sn.

We evaluate the spin-singlet and spin-triplet pair condensations with
the following pairing components:
\begin{align}
S_{\rho_n} &= \int d\bm{r} |\tilde{\rho}_t(\bm{r})|^2, \\
S_{J_n} &= R^2\int d\bm{r} |\tilde{\bm{J}}_t(\bm{r})|^2.
\end{align}
They have exactly the same local density dependence that appears in the pair energy. 
The constant $R^2=10$ fm$^2$ is estimated from the 
ratio $|\tilde{C}_t^J/\tilde{C}_t^\rho|$ of the G3RS potential and is 
introduced to make the units of the two quantities identical.

In Fig. \ref{fig:alice},
the results from neutron pair densities in the Ca and Sn isotopes are presented.
The spin-singlet component $S_{\rho_n}$ shows finite values except in neutron closed-shell nuclei.
Even though the attractive pair EDF is present only in the spin-singlet channel, our results show nonzero values for the spin-triplet component $S_{J_n}$, namely, a coexistence of the spin-singlet and spin-triplet condensates is suggested in finite nuclei.
This is also expected in 
Eqs.~(\ref{eq:rhot_sph}) and (\ref{eq:Jt_sph}). 
A similar feature has been discussed in 
the case of the $T=1$ proton-neutron pair~\cite{PhysRevC.90.054332}, in
condensed matter~\cite{PhysRevLett.87.037004}, and in ultracold Fermi gas~\cite{  PhysRevLett.107.195304,PhysRevB.84.014512,PhysRevA.92.013615} in the 
presence of the spin-orbit ($\boldsymbol{k}\otimes \boldsymbol{\sigma}$ type) interaction.
Notice that a direct comparison of $S_{\rho_n}$ and $S_{J_n}$ does not make sense as their relative value depends on the introduced constant $R^2$.
However, the isotopic dependence indicates that the spin-triplet pairing is more sensitive to the shell orbits involved than the spin-singlet one is.
$S_{\rho_n}$ is enhanced in the midshell region with high degeneracy, such as 
in the $f_{7/2}$ and $f_{5/2}$ orbits in Ca isotopes
and $50<N<82$ and $82<N<126$ in Sn isotopes,
while $S_{J_n}$ shows a stronger orbital dependence; 
we see an enhancement (a reduction) in $S_{J_n}$ in the isotope where the neutron
Fermi energy is around $j_{>}$ ($j_<$) orbit in $f_{7/2}$ and $f_{5/2}$ in Ca isotopes
and an enhancement in the intruder region in the middle shell
in Sn isotopes.

\begin{figure}
\begin{center}
\includegraphics[width=80mm]{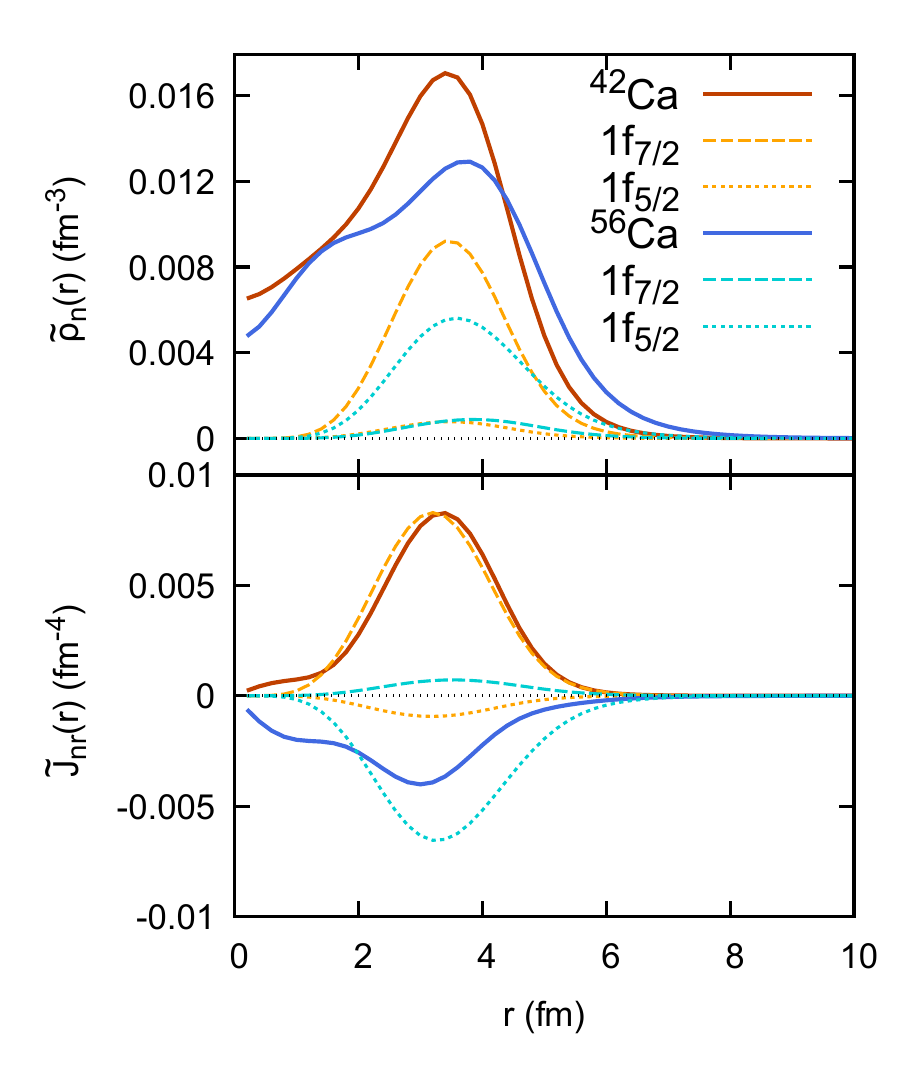}
\end{center}
\caption{
The neutron local pair density $\tilde{\rho}_n(r)$ and 
the radial component of the neutron tensor pair density 
$\tilde{J}_{nr}(r)$ of $^{42}$Ca and $^{56}$Ca and the contributions from $1f_{7/2}$ and $1f_{5/2}$ orbits.}
\label{fig:davide}
\end{figure}

To analyze the contributions from the $j_>$ and $j_<$ orbits,
we take $^{42}$Ca and $^{56}$Ca as representative cases,
where two particles are supposed to occupy the $f_{7/2}$ and $f_{5/2}$ orbits mainly.
The pair density distributions
$\tilde{\rho}_n(r)$ and $\tilde{J}_{nr}(r)$
together with the contributions from $f_{7/2}$ and $f_{5/2}$ orbits are plotted in Fig. \ref{fig:davide}. 
Both the spin-singlet and spin-triplet neutron pair densities have finite values
in the $^{42}$Ca and $^{56}$Ca nuclei.
There, the $f_{7/2}$ and $f_{5/2}$ neutrons have dominant contributions as expected.
For the spin-singlet density,
they have a coherent contribution,
and the total pair densities are composed of the coherent addition from the other orbits as well (not shown in the figure),
whereas the neutrons in the $f_{7/2}$ and $f_{5/2}$ orbits
contribute 
in a destructive way to the spin-triplet density.
The dominant contribution for $\tilde{J}_{nr}(r)$ in $^{42}$Ca
 is from the $f_{7/2}$ orbit, while $\tilde{J}_{nr}(r)$ in $^{56}$Ca is composed of the multiple orbits including those not shown in the figure. This indicates the magicity at $N=34$
is weaker than that at $N=28$.

\subsection{Spin-triplet pair EDF}
In place of the spin-singlet pair EDF, we employ
 the spin-triplet pair EDF.
The coupling constant of the spin-triplet pair EDF is 
adjusted to reproduce the pairing energy 
of $^{44}$Ca obtained in the spin-singlet pair EDF
($\tilde{C}_n^{J1} = -46.125$ MeV fm$^5$). 
The pairing energy is $-5.08$ and $-5.25$ MeV in the singlet-pair EDF and triplet-pair EDF, as shown in Figs.~\ref{fig:singlet-triplet-gap}(c) 
and \ref{fig:singlet-triplet-gap}(d).

Figure \ref{fig:singlet-triplet} shows the 
spin-singlet and spin-triplet pairing components
calculated with either the 
spin-singlet or the spin-triplet pair EDF for the Ca isotopes.
The spin-singlet pairing component calculated with the 
spin-singlet pair EDF and the spin-triplet pairing component calculated with the spin-triplet pair EDF 
have very similar properties: one sees the 
collapse of the pairing at the magic numbers $N=20, 28, 32$, and $40$ (there is a tiny difference in $N=34$)
and the neutron-number dependence of the relative size of the pairing component.
We also note that there is little difference in the pairing energy, the chemical potential, and other observables of the particle-hole type.
The coupling constant of the spin-triplet pair EDF $\tilde{C}^{J1}_t$
can be related to the Skyrme parameters as
$\tilde{C}_t^{J1} = [t_2(1+x_2)+5t_o + 2W_0)]/8$ and 
are repulsive for many Skyrme interactions
such as SIII (18.125 MeV fm$^5$) \cite{Beiner197529}, SLy4 (30.75 MeV fm$^5$), SLy5 (31.5 MeV fm$^5$) \cite{Chabanat1998231},
and SkP (5.486 MeV fm$^5$),
but can be attractive for the Skyrme interactions that include
the tensor interaction such as SLy5 + T ($-53.5$ MeV fm$^5$) \cite{COLO2007227}, and
14 Skyrme parameters out of 36 T$IJ$ parameter sets in Ref.~\cite{PhysRevC.76.014312}.
Although the coupling constants of the EDF can be taken
arbitrarily in the framework of the nuclear density functional theory, 
the tensor interaction will have a large impact on the 
property of the spin-triplet pairing coupling constant.

\subsection{Roles of the spin-orbit EDF}
\begin{figure}
\includegraphics[width=86mm]{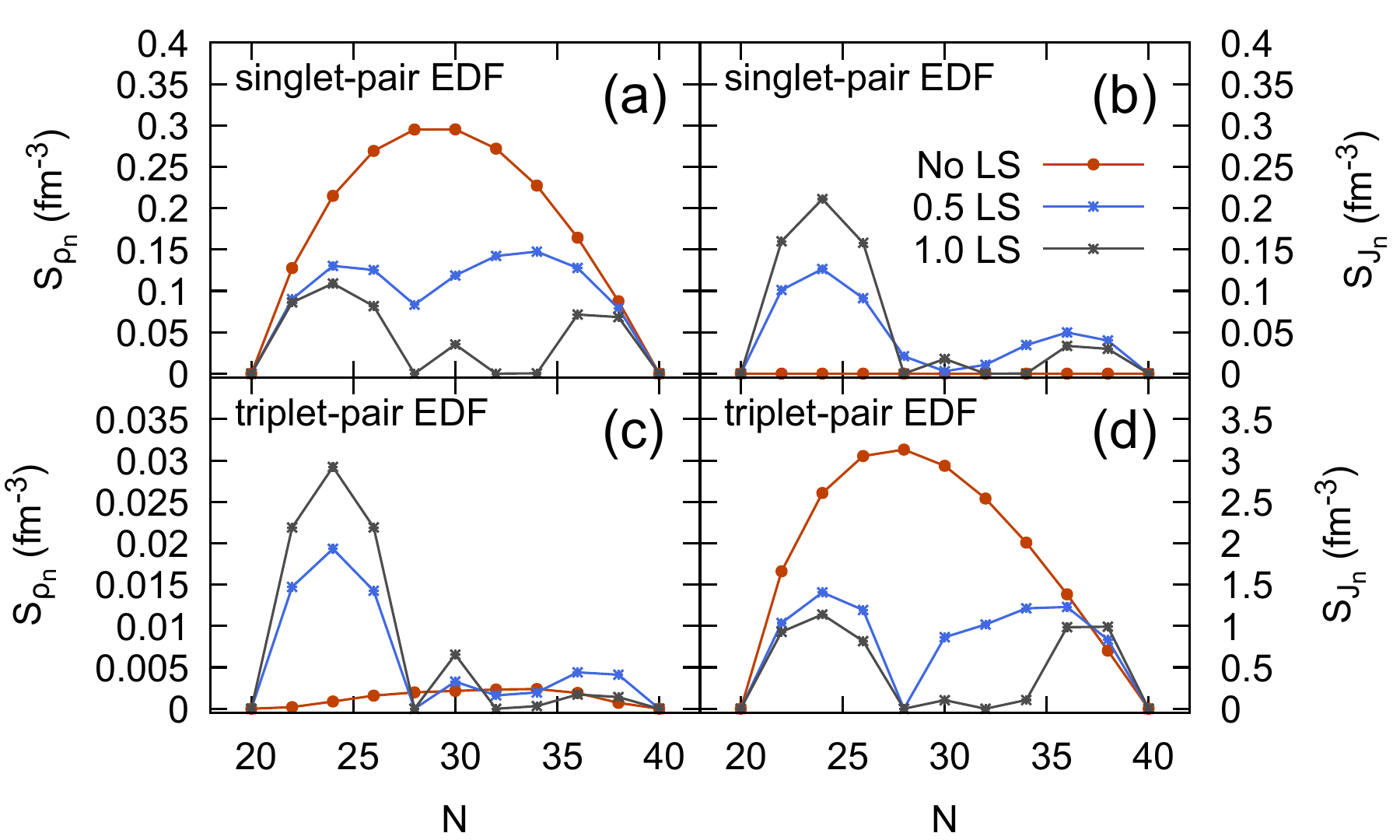} 
\caption{Spin-singlet and spin-triplet pairing components, $S_{\rho_n}$ and $S_{J_n}$, calculated in the Ca isotopes by changing the coupling constant of the spin-orbit EDF. The labels 1.0~LS, 0.5~LS, and No LS represent the calculations with
the original spin-orbit EDF, the reduced one multiplied by a factor of 0.5, and the calculation without the spin-orbit EDF, respectively.
}
\label{fig:singlet-triplet}
\end{figure}

The spin-orbit splitting is expected to play an 
important role in the spin-triplet pairing as anticipated from the expression of the tensor pair density (\ref{eq:Jt_sph}).
In Fig.~\ref{fig:singlet-triplet}, we also present the pairing components
$S_{\rho_n}$ and $S_{J_n}$ obtained by changing the coupling constant of the spin-orbit EDF while keeping the pairing coupling constants to the original values.
Three spin-orbit EDFs are considered: the original spin-orbit EDF (1.0 LS), the reduced one multiplied by a factor of 0.5 (0.5 LS), and no spin-orbit EDF (No LS).
First, suppose that the spin-orbit EDF is neglected.
Then, the spin-singlet pair EDF
promotes only the spin-singlet pair condensate, and the spin-triplet pairing component $S_{J_n}$ is zero [Figs.~\ref{fig:singlet-triplet}(a) and \ref{fig:singlet-triplet}(b)].
One is tempted to the opposite conclusion when the spin-triplet pair EDF is considered.
However, 
the appearance of the spin-triplet pairing component inevitably induces the spin-orbit splitting and the spin-singlet pairing component. 
This results in a nonzero spin-singlet pairing component,
although the induced amount is tiny.
As a result,
commonly with the spin-singlet and spin-triplet EDFs, the corresponding pairing component takes the maximum at around $N=28$, which is around the half-filled situation of the 14-fold degenerated $f$ orbit [Figs.~\ref{fig:singlet-triplet}(a) and \ref{fig:singlet-triplet}(d)]. The pairing component becomes zero at LS-closed shells $N=20$ and $N=40$.

The behavior of the pairing components $S_{\rho_n}$ and $S_{J_n}$ shows a drastic change with the value of the coupling constant of the spin-orbit EDF, although we do not change the pair EDF itself.
The spin-orbit EDF decreases the pairing component due to the lower degeneracy of the single-particle levels
[Figs.~\ref{fig:singlet-triplet}(a) and \ref{fig:singlet-triplet}(d)], but it enhances the coexistence of $S_{\rho_n}$ and $S_{J_n}$ [Figs.~\ref{fig:singlet-triplet}(b) and \ref{fig:singlet-triplet}(c)].
\footnote{The mixture of the spin-singlet (even-parity) and spin-triplet (odd-parity) components due to the spin-orbit field was discussed in the proton-neutron channel within the three-body model for $^{16}$O$+p+n$ \cite{PhysRevC.90.054332}. }
By comparing the $0.5$ LS and $1.0$ LS cases,
one can see an enhancement of the mixing 
in the region $20<N<28$ and the suppression in the neutron-rich side with $N>30$. The suppression is common for the main pairing component and the induced pairing component due to the lesser degeneracy of the single-particle levels with increasing the coupling constant of the spin-orbit EDF.

\begin{figure}
\includegraphics[width=86mm]{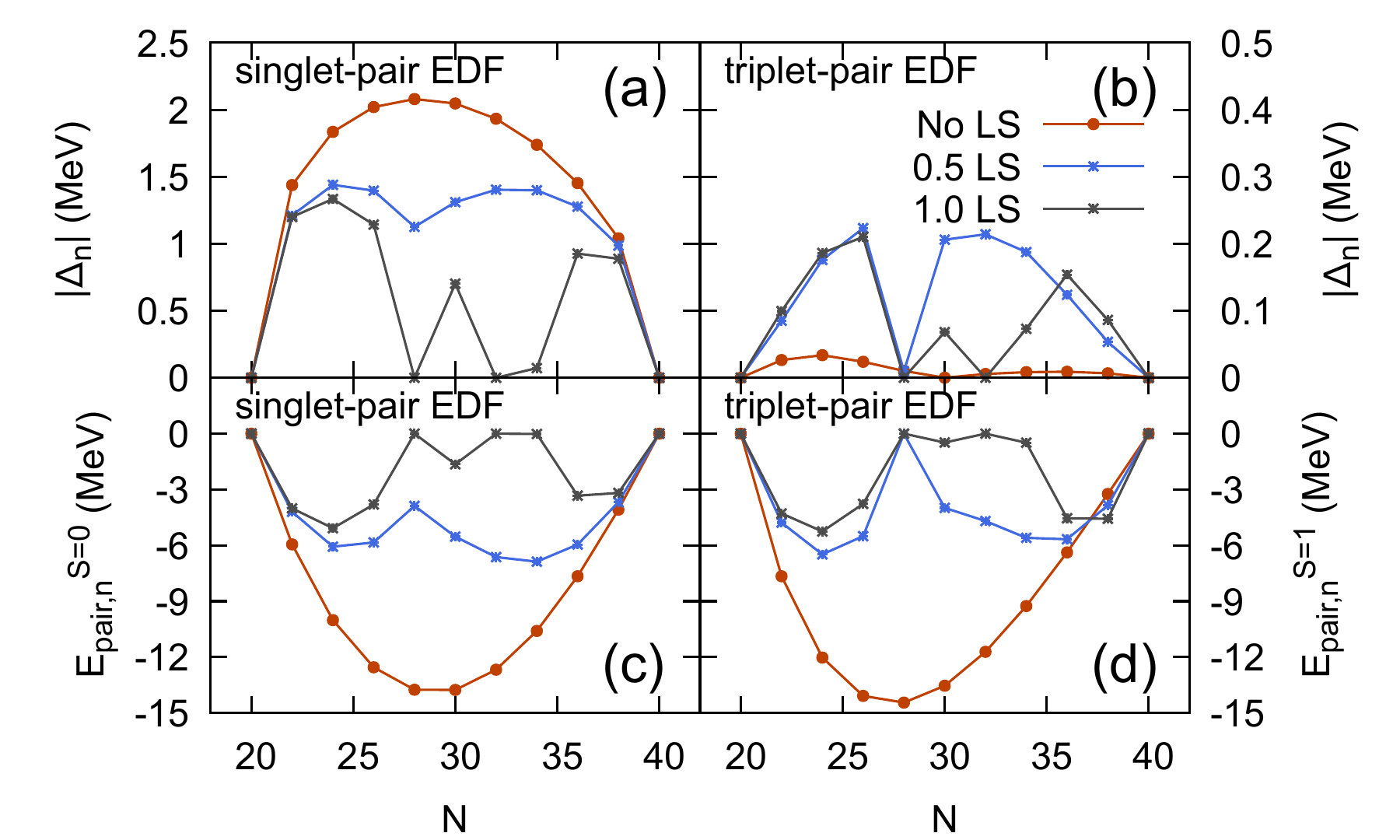} 
\caption{
Pairing gap $|\Delta_n|$ and pairing energy $E^{S=0,1}_{{\rm pair},n}$ calculated for the singlet- and triplet-pair EDF 
in the Ca isotopes by changing the coupling constant of the spin-orbit EDF.}
\label{fig:singlet-triplet-gap}
\end{figure}

We also plot the pairing gap and the pairing energy in Fig.~\ref{fig:singlet-triplet-gap}. Figures~\ref{fig:singlet-triplet}(a) and \ref{fig:singlet-triplet-gap}(a) show that the singlet-pairing component $S_{\rho_n}$ and the pairing gap $\Delta_n$ behave in a similar way in the case of singlet-pair EDF, 
while the triplet-pairing component $S_{J_n}$ [Fig.~\ref{fig:singlet-triplet}(d)] and the pairing gap [Fig.~\ref{fig:singlet-triplet-gap}(b)] in the case of the triplet-pair EDF behave in a different way.
A strong reduction of the averaged gap for the triplet pairing in the ``No LS" case [Fig.~\ref{fig:singlet-triplet-gap}(b)] is due to a low tensor density ${\sf J}$.
The pairing energies for the singlet-pair EDF [Fig.~\ref{fig:singlet-triplet-gap}(c)] and the triplet-pair EDF [Fig.~\ref{fig:singlet-triplet-gap}(d)] take similar values as a function of the neutron number, although the coupling constants of the two pair EDFs are adjusted only at $N=24$. 
The agreement of the pairing energy shows that the triplet-pair EDF can include a contribution similar to that of the singlet-pair EDF.

\subsection{Relevant observables}

The small values of the pairing gap defined by Eq.~(\ref{eq:gap}) may not correspond to the experimental OES
for the triplet-pair EDF [Fig.~\ref{fig:singlet-triplet-gap}(b)],
similar to when the singlet-pair Hamiltonian contains derivative terms \cite{0954-3899-45-2-024004}.
We note that another average gap in which $\phi_2^{(t)\ast}$ is replaced by $\phi_1^{(t)\ast}$ in Eq.~(\ref{eq:gap}) behaves even worse in the case of the triplet-pair EDF, because of the singlet-pair amplitude in the denominator that is very small as expected from Fig. \ref{fig:singlet-triplet}(c).

\begin{figure}[H]
\includegraphics[width=80mm]{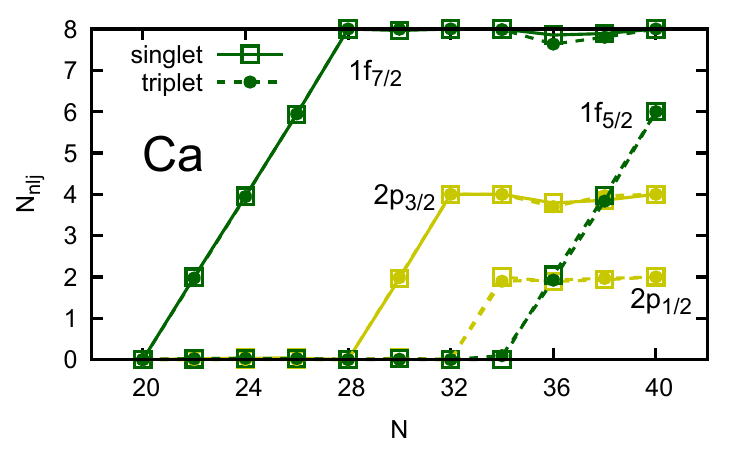}
\caption{
Occupation numbers of $pf$-shell orbits in Ca isotopes calculated with the 
singlet- and triplet-pair EDFs.
\label{fig:occupation}
}
\end{figure}

To analyze the influence of the type of pairing in the occupation, we plot the occupation of each neutron orbit in Fig.~\ref{fig:occupation}.
The occupation number increases in the order of the single-particle energies with the neutron number increases, showing that the 
dominant pair correlation is within a single orbit near the Fermi energy.
There is no significant difference between the occupation numbers calculated with the singlet- and triplet-pair EDFs.
The difference is barely visible  around $34\le N \le 38$. 

\begin{figure}
\includegraphics[width=86mm]{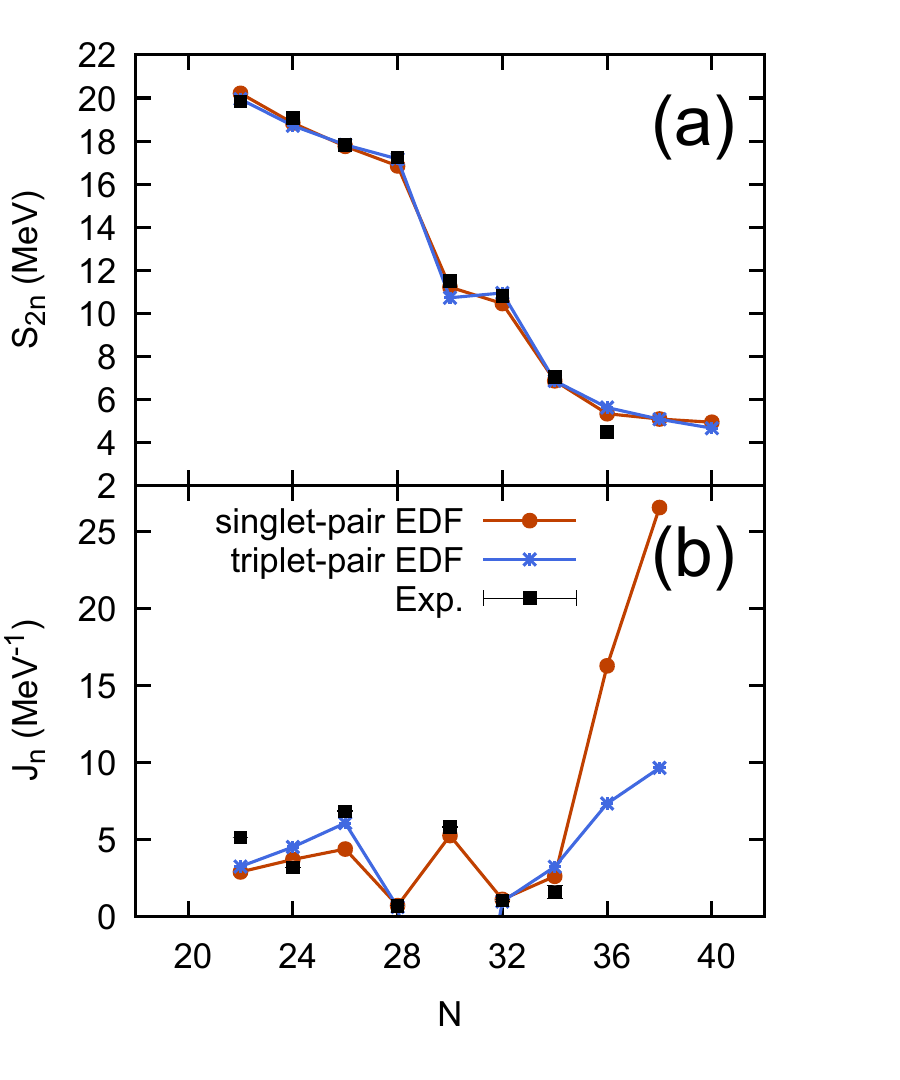} 
\caption{
Two-neutron separation energies and moments of inertia of the neutron pairing rotation in the Ca isotopes calculated for the singlet- and triplet-pair EDFs. The inertia calculated with the triplet-pair EDF at $N=30$ ($-17.86$ MeV$^{-1}$) is not shown in the figure.
\label{fig:prmoi}
}
\end{figure}

The pairing rotational moment of inertia can be 
a relevant observable of the pairing condensation \cite{PhysRevLett.116.152502} even in the presence of the triplet-pair EDF. 
Figure~\ref{fig:prmoi} shows the 
two-neutron separation energies $S_{2n}(N) = E(N-2)-E(N)$ and 
the neutron pairing rotational moments of inertia 
calculated from the three HFB energy differences ${\cal J}_n(N) = 4/\Delta S_{2n}(N)
= 4/[ E(N+2)-2E(N)+E(N-2)]$.
Unlike the relation between the pairing gap and the OES, the pairing rotational moment of inertia holds a good correspondence with the 
experimental values of the double binding-energy differences.
The moment of inertia at $N=30$ becomes negative ($-17.86$ MeV$^{-1}$) for the triplet-pair EDF, while the singlet-pair EDF reproduces the experimental value.
The negative value originates from the staggering behavior 
of $S_{2n}$ at $N=30$ seen in Fig.~\ref{fig:prmoi} (a).
Two of the HFB states ($N=28$ and 32) used to computed the inertia at $N=30$ are in the normal states corresponding to the  $f_{7/2}$ and $p_{3/2}$ shell closures, and 
the inertia at $N=30$ does not correspond to the pairing indicator, as the expression of the inertia based on the double binding-energy differences assumes that the three nearby isotopes have similar pairing structures, and this assumption does not hold in this case.
We can see a difference in the moment of inertia in the neutron-rich region at $N=36$ and 38 for the singlet-pair and triplet-pair EDF.
The binding energies in the neutron-rich isotopes may 
determine the coupling constant of the triplet-pair EDF.

\section{Summary}
We analyzed the spin-triplet pair condensation of like particles
in singly closed nuclei. The relevant quantity 
to the spin-triplet pair condensation is the nine-component spin-current pair density. 
One component can be finite in the HFB calculation within the spherical symmetry.
We have demonstrated that
the spin-singlet and spin-triplet pairing condensates coexist in open-shell nuclei, and one component of the pair EDF can induce the other component.
The spin-orbit splitting is shown to play an essential role in the coexistence
of the two types of pair condensates,
because the spin is no longer a good quantum number.

The inclusion of both the spin-singlet and spin-triplet pair EDFs into
the nuclear EDF will enable us more detailed description of the 
nuclear pairing condensation, including the isotope and isotone dependence,
and deepen the understanding of the origin of the spin-orbit splitting and the role of the tensor force in the pairing channel.
In describing open-shell nuclei with deformation, 
the pseudoscalar and pseudotensor components should be considered. 
In subsequent works, we will present such developments.

\section*{Acknowledgments}
The authors express their sincere gratitude to T.~Naito, W.~Nazarewicz, H.~Tajima, and Y.~Yanase for their conscientious review of the manuscript and invaluable insights. Additionally, the authors acknowledge the invaluable contributions of members of the PHANES Collaboration (M.~Dozono, M.~Matsuo, S.~Ota, and S.~Shimoura) for their insightful discussions.
This work was supported by the 
JSPS KAKENHI (Grants No. JP19K03824, No. JP19K03872, No. JP19KK0343, and No. JP20K03964).

\bibliography{toddpair}

\begin{thebibliography}{53}%
\makeatletter
\providecommand \@ifxundefined [1]{%
 \@ifx{#1\undefined}
}%
\providecommand \@ifnum [1]{%
 \ifnum #1\expandafter \@firstoftwo
 \else \expandafter \@secondoftwo
 \fi
}%
\providecommand \@ifx [1]{%
 \ifx #1\expandafter \@firstoftwo
 \else \expandafter \@secondoftwo
 \fi
}%
\providecommand \natexlab [1]{#1}%
\providecommand \enquote  [1]{``#1''}%
\providecommand \bibnamefont  [1]{#1}%
\providecommand \bibfnamefont [1]{#1}%
\providecommand \citenamefont [1]{#1}%
\providecommand \href@noop [0]{\@secondoftwo}%
\providecommand \href [0]{\begingroup \@sanitize@url \@href}%
\providecommand \@href[1]{\@@startlink{#1}\@@href}%
\providecommand \@@href[1]{\endgroup#1\@@endlink}%
\providecommand \@sanitize@url [0]{\catcode `\\12\catcode `\$12\catcode
  `\&12\catcode `\#12\catcode `\^12\catcode `\_12\catcode `\%12\relax}%
\providecommand \@@startlink[1]{}%
\providecommand \@@endlink[0]{}%
\providecommand \url  [0]{\begingroup\@sanitize@url \@url }%
\providecommand \@url [1]{\endgroup\@href {#1}{\urlprefix }}%
\providecommand \urlprefix  [0]{URL }%
\providecommand \Eprint [0]{\href }%
\providecommand \doibase [0]{https://doi.org/}%
\providecommand \selectlanguage [0]{\@gobble}%
\providecommand \bibinfo  [0]{\@secondoftwo}%
\providecommand \bibfield  [0]{\@secondoftwo}%
\providecommand \translation [1]{[#1]}%
\providecommand \BibitemOpen [0]{}%
\providecommand \bibitemStop [0]{}%
\providecommand \bibitemNoStop [0]{.\EOS\space}%
\providecommand \EOS [0]{\spacefactor3000\relax}%
\providecommand \BibitemShut  [1]{\csname bibitem#1\endcsname}%
\let\auto@bib@innerbib\@empty
\bibitem [{\citenamefont {Dean}\ and\ \citenamefont
  {Hjorth-Jensen}(2003)}]{RevModPhys.75.607}%
  \BibitemOpen
  \bibfield  {author} {\bibinfo {author} {\bibfnamefont {D.~J.}\ \bibnamefont
  {Dean}}\ and\ \bibinfo {author} {\bibfnamefont {M.}~\bibnamefont
  {Hjorth-Jensen}},\ }\bibfield  {title} {\bibinfo {title} {{Pairing in nuclear
  systems: from neutron stars to finite nuclei}},\ }\href
  {https://doi.org/10.1103/RevModPhys.75.607} {\bibfield  {journal} {\bibinfo
  {journal} {Rev. Mod. Phys.}\ }\textbf {\bibinfo {volume} {75}},\ \bibinfo
  {pages} {607} (\bibinfo {year} {2003})}\BibitemShut {NoStop}%
\bibitem [{\citenamefont {Brink}\ and\ \citenamefont {Broglia}(2005)}]{05BB}%
  \BibitemOpen
  \bibfield  {author} {\bibinfo {author} {\bibfnamefont {D.~M.}\ \bibnamefont
  {Brink}}\ and\ \bibinfo {author} {\bibfnamefont {R.~A.}\ \bibnamefont
  {Broglia}},\ }\href@noop {} {\emph {\bibinfo {title} {Nuclear Superfluidity:
  Pairing in Finite Systems}}},\ Cambridge Monographs on Particle Physics,
  Nuclear Physics and Cosmology\ (\bibinfo  {publisher} {Cambridge University
  Press},\ \bibinfo {address} {Cambridge, England},\ \bibinfo {year}
  {2005})\BibitemShut {NoStop}%
\bibitem [{\citenamefont {Bardeen}\ \emph {et~al.}(1957)\citenamefont
  {Bardeen}, \citenamefont {Cooper},\ and\ \citenamefont
  {Schrieffer}}]{57Bardeen}%
  \BibitemOpen
  \bibfield  {author} {\bibinfo {author} {\bibfnamefont {J.}~\bibnamefont
  {Bardeen}}, \bibinfo {author} {\bibfnamefont {L.~N.}\ \bibnamefont
  {Cooper}},\ and\ \bibinfo {author} {\bibfnamefont {J.~R.}\ \bibnamefont
  {Schrieffer}},\ }\bibfield  {title} {\bibinfo {title} {{Theory of
  superconductivity}},\ }\href {https://doi.org/10.1103/PhysRev.108.1175}
  {\bibfield  {journal} {\bibinfo  {journal} {Phys. Rev.}\ }\textbf {\bibinfo
  {volume} {108}},\ \bibinfo {pages} {1175} (\bibinfo {year}
  {1957})}\BibitemShut {NoStop}%
\bibitem [{\citenamefont {Sigrist}\ and\ \citenamefont
  {Ueda}(1991)}]{1991Sigrist_rev}%
  \BibitemOpen
  \bibfield  {author} {\bibinfo {author} {\bibfnamefont {M.}~\bibnamefont
  {Sigrist}}\ and\ \bibinfo {author} {\bibfnamefont {K.}~\bibnamefont {Ueda}},\
  }\bibfield  {title} {\bibinfo {title} {Phenomenological theory of
  unconventional superconductivity},\ }\href
  {https://doi.org/10.1103/RevModPhys.63.239} {\bibfield  {journal} {\bibinfo
  {journal} {Rev. Mod. Phys.}\ }\textbf {\bibinfo {volume} {63}},\ \bibinfo
  {pages} {239} (\bibinfo {year} {1991})}\BibitemShut {NoStop}%
\bibitem [{\citenamefont {Takatsuka}\ and\ \citenamefont
  {Tamagaki}(1993)}]{PTPS.112.27}%
  \BibitemOpen
  \bibfield  {author} {\bibinfo {author} {\bibfnamefont {T.}~\bibnamefont
  {Takatsuka}}\ and\ \bibinfo {author} {\bibfnamefont {R.}~\bibnamefont
  {Tamagaki}},\ }\bibfield  {title} {\bibinfo {title} {{Superfluidity in
  neutron star matter and symmetric nuclear matter}},\ }\href
  {https://doi.org/10.1143/PTP.112.27} {\bibfield  {journal} {\bibinfo
  {journal} {Prog. Theor. Phys. Suppl.}\ }\textbf {\bibinfo {volume} {112}},\
  \bibinfo {pages} {27} (\bibinfo {year} {1993})}\BibitemShut {NoStop}%
\bibitem [{\citenamefont {Mackenzie}\ and\ \citenamefont
  {Maeno}(2003)}]{2003Mack_rev}%
  \BibitemOpen
  \bibfield  {author} {\bibinfo {author} {\bibfnamefont {A.~P.}\ \bibnamefont
  {Mackenzie}}\ and\ \bibinfo {author} {\bibfnamefont {Y.}~\bibnamefont
  {Maeno}},\ }\bibfield  {title} {\bibinfo {title} {{The superconductivity of
  ${\mathrm{Sr}}_{2}{\mathrm{RuO}}_{4}$ and the physics of spin-triplet
  pairing}},\ }\href {https://doi.org/10.1103/RevModPhys.75.657} {\bibfield
  {journal} {\bibinfo  {journal} {Rev. Mod. Phys.}\ }\textbf {\bibinfo {volume}
  {75}},\ \bibinfo {pages} {657} (\bibinfo {year} {2003})}\BibitemShut
  {NoStop}%
\bibitem [{\citenamefont {Frauendorf}\ and\ \citenamefont
  {Macchiavelli}(2014)}]{Frauendorf:2014mja}%
  \BibitemOpen
  \bibfield  {author} {\bibinfo {author} {\bibfnamefont {S.}~\bibnamefont
  {Frauendorf}}\ and\ \bibinfo {author} {\bibfnamefont {A.~O.}\ \bibnamefont
  {Macchiavelli}},\ }\bibfield  {title} {\bibinfo {title} {{Overview of
  neutron\textendash{}proton pairing}},\ }\href
  {https://doi.org/10.1016/j.ppnp.2014.07.001} {\bibfield  {journal} {\bibinfo
  {journal} {Prog. Part. Nucl. Phys.}\ }\textbf {\bibinfo {volume} {78}},\
  \bibinfo {pages} {24} (\bibinfo {year} {2014})}\BibitemShut {NoStop}%
\bibitem [{\citenamefont {Osheroff}\ \emph
  {et~al.}(1972{\natexlab{a}})\citenamefont {Osheroff}, \citenamefont
  {Richardson},\ and\ \citenamefont {Lee}}]{1972Osheroff_01}%
  \BibitemOpen
  \bibfield  {author} {\bibinfo {author} {\bibfnamefont {D.~D.}\ \bibnamefont
  {Osheroff}}, \bibinfo {author} {\bibfnamefont {R.~C.}\ \bibnamefont
  {Richardson}},\ and\ \bibinfo {author} {\bibfnamefont {D.~M.}\ \bibnamefont
  {Lee}},\ }\bibfield  {title} {\bibinfo {title} {{Evidence for a new phase of
  solid ${\mathrm{He}}^{3}$}},\ }\href
  {https://doi.org/10.1103/PhysRevLett.28.885} {\bibfield  {journal} {\bibinfo
  {journal} {Phys. Rev. Lett.}\ }\textbf {\bibinfo {volume} {28}},\ \bibinfo
  {pages} {885} (\bibinfo {year} {1972}{\natexlab{a}})}\BibitemShut {NoStop}%
\bibitem [{\citenamefont {Osheroff}\ \emph
  {et~al.}(1972{\natexlab{b}})\citenamefont {Osheroff}, \citenamefont {Gully},
  \citenamefont {Richardson},\ and\ \citenamefont {Lee}}]{1972Osheroff_02}%
  \BibitemOpen
  \bibfield  {author} {\bibinfo {author} {\bibfnamefont {D.~D.}\ \bibnamefont
  {Osheroff}}, \bibinfo {author} {\bibfnamefont {W.~J.}\ \bibnamefont {Gully}},
  \bibinfo {author} {\bibfnamefont {R.~C.}\ \bibnamefont {Richardson}},\ and\
  \bibinfo {author} {\bibfnamefont {D.~M.}\ \bibnamefont {Lee}},\ }\bibfield
  {title} {\bibinfo {title} {{New magnetic phenomena in liquid
  ${\mathrm{He}}^{3}$ below 3 mK}},\ }\href
  {https://doi.org/10.1103/PhysRevLett.29.920} {\bibfield  {journal} {\bibinfo
  {journal} {Phys. Rev. Lett.}\ }\textbf {\bibinfo {volume} {29}},\ \bibinfo
  {pages} {920} (\bibinfo {year} {1972}{\natexlab{b}})}\BibitemShut {NoStop}%
\bibitem [{\citenamefont {Leggett}(1972)}]{1972Leggett}%
  \BibitemOpen
  \bibfield  {author} {\bibinfo {author} {\bibfnamefont {A.~J.}\ \bibnamefont
  {Leggett}},\ }\bibfield  {title} {\bibinfo {title} {{Interpretation of recent
  results on ${\mathrm{He}}^{3}$ below 3 mK: A new liquid phase?}},\ }\href
  {https://doi.org/10.1103/PhysRevLett.29.1227} {\bibfield  {journal} {\bibinfo
   {journal} {Phys. Rev. Lett.}\ }\textbf {\bibinfo {volume} {29}},\ \bibinfo
  {pages} {1227} (\bibinfo {year} {1972})}\BibitemShut {NoStop}%
\bibitem [{\citenamefont {Fulde}\ and\ \citenamefont
  {Ferrell}(1964)}]{1964Fulde}%
  \BibitemOpen
  \bibfield  {author} {\bibinfo {author} {\bibfnamefont {P.}~\bibnamefont
  {Fulde}}\ and\ \bibinfo {author} {\bibfnamefont {R.~A.}\ \bibnamefont
  {Ferrell}},\ }\bibfield  {title} {\bibinfo {title} {{Superconductivity in a
  strong spin-exchange field}},\ }\href
  {https://doi.org/10.1103/PhysRev.135.A550} {\bibfield  {journal} {\bibinfo
  {journal} {Phys. Rev.}\ }\textbf {\bibinfo {volume} {135}},\ \bibinfo {pages}
  {A550} (\bibinfo {year} {1964})}\BibitemShut {NoStop}%
\bibitem [{\citenamefont {Larkin}\ and\ \citenamefont
  {Ovchinnikov}(1964)}]{1964Larkin}%
  \BibitemOpen
  \bibfield  {author} {\bibinfo {author} {\bibfnamefont {A.~I.}\ \bibnamefont
  {Larkin}}\ and\ \bibinfo {author} {\bibfnamefont {Y.~N.}\ \bibnamefont
  {Ovchinnikov}},\ }\bibfield  {title} {\bibinfo {title} {Nonuniform state of
  superconductors},\ }\href@noop {} {\bibfield  {journal} {\bibinfo  {journal}
  {Zh. Eksp. Teor. Fiz.}\ }\textbf {\bibinfo {volume} {47}},\ \bibinfo {pages}
  {1136} (\bibinfo {year} {1964})}\BibitemShut {NoStop}%
\bibitem [{\citenamefont {Casalbuoni}\ and\ \citenamefont
  {Nardulli}(2004)}]{2004Casal}%
  \BibitemOpen
  \bibfield  {author} {\bibinfo {author} {\bibfnamefont {R.}~\bibnamefont
  {Casalbuoni}}\ and\ \bibinfo {author} {\bibfnamefont {G.}~\bibnamefont
  {Nardulli}},\ }\bibfield  {title} {\bibinfo {title} {{Inhomogeneous
  superconductivity in condensed matter and QCD}},\ }\href
  {https://doi.org/10.1103/RevModPhys.76.263} {\bibfield  {journal} {\bibinfo
  {journal} {Rev. Mod. Phys.}\ }\textbf {\bibinfo {volume} {76}},\ \bibinfo
  {pages} {263} (\bibinfo {year} {2004})}\BibitemShut {NoStop}%
\bibitem [{\citenamefont {Yanase}(2009)}]{2009Yanase}%
  \BibitemOpen
  \bibfield  {author} {\bibinfo {author} {\bibfnamefont {Y.}~\bibnamefont
  {Yanase}},\ }\bibfield  {title} {\bibinfo {title} {{Angular
  Fulde-Ferrell-Larkin-Ovchinnikov state in cold fermion gases in a toroidal
  trap}},\ }\href {https://doi.org/10.1103/PhysRevB.80.220510} {\bibfield
  {journal} {\bibinfo  {journal} {Phys. Rev. B}\ }\textbf {\bibinfo {volume}
  {80}},\ \bibinfo {pages} {220510(R)} (\bibinfo {year} {2009})}\BibitemShut
  {NoStop}%
\bibitem [{\citenamefont {Bergeret}\ and\ \citenamefont
  {Volkov}(2023)}]{2023Berg}%
  \BibitemOpen
  \bibfield  {author} {\bibinfo {author} {\bibfnamefont {F.~S.}\ \bibnamefont
  {Bergeret}}\ and\ \bibinfo {author} {\bibfnamefont {A.~F.}\ \bibnamefont
  {Volkov}},\ }\bibfield  {title} {\bibinfo {title} {Triplet odd-frequency
  superconductivity in hybrid superconductor–ferromagnet structures},\ }\href
  {https://doi.org/https://doi.org/10.1016/j.aop.2023.169232} {\bibfield
  {journal} {\bibinfo  {journal} {Ann. Phys.}\ }\textbf {\bibinfo {volume}
  {456}},\ \bibinfo {pages} {169232} (\bibinfo {year} {2023})}\BibitemShut
  {NoStop}%
\bibitem [{\citenamefont {Aoki}\ \emph {et~al.}(2019)\citenamefont {Aoki},
  \citenamefont {Ishida},\ and\ \citenamefont {Flouquet}}]{2019Aoki_rev}%
  \BibitemOpen
  \bibfield  {author} {\bibinfo {author} {\bibfnamefont {D.}~\bibnamefont
  {Aoki}}, \bibinfo {author} {\bibfnamefont {K.}~\bibnamefont {Ishida}},\ and\
  \bibinfo {author} {\bibfnamefont {J.}~\bibnamefont {Flouquet}},\ }\bibfield
  {title} {\bibinfo {title} {{Review of U-based ferromagnetic superconductors:
  Comparison between UGe2, URhGe, and UCoGe}},\ }\href
  {https://doi.org/10.7566/JPSJ.88.022001} {\bibfield  {journal} {\bibinfo
  {journal} {J. Phys. Soc. Jpn}\ }\textbf {\bibinfo {volume} {88}},\ \bibinfo
  {pages} {022001} (\bibinfo {year} {2019})}\BibitemShut {NoStop}%
\bibitem [{\citenamefont {Cai}\ \emph {et~al.}(2020)\citenamefont {Cai},
  \citenamefont {Sun}, \citenamefont {Xia}, \citenamefont {Wu}, \citenamefont
  {Liu}, \citenamefont {Liu}, \citenamefont {Gong}, \citenamefont {Yao},
  \citenamefont {Guo},\ and\ \citenamefont {Wang}}]{2020Cai}%
  \BibitemOpen
  \bibfield  {author} {\bibinfo {author} {\bibfnamefont {W.}~\bibnamefont
  {Cai}}, \bibinfo {author} {\bibfnamefont {H.}~\bibnamefont {Sun}}, \bibinfo
  {author} {\bibfnamefont {W.}~\bibnamefont {Xia}}, \bibinfo {author}
  {\bibfnamefont {C.}~\bibnamefont {Wu}}, \bibinfo {author} {\bibfnamefont
  {Y.}~\bibnamefont {Liu}}, \bibinfo {author} {\bibfnamefont {H.}~\bibnamefont
  {Liu}}, \bibinfo {author} {\bibfnamefont {Y.}~\bibnamefont {Gong}}, \bibinfo
  {author} {\bibfnamefont {D.-X.}\ \bibnamefont {Yao}}, \bibinfo {author}
  {\bibfnamefont {Y.}~\bibnamefont {Guo}},\ and\ \bibinfo {author}
  {\bibfnamefont {M.}~\bibnamefont {Wang}},\ }\bibfield  {title} {\bibinfo
  {title} {{Pressure-induced superconductivity and structural transition in
  ferromagnetic ${\mathrm{CrSiTe}}_{3}$}},\ }\href
  {https://doi.org/10.1103/PhysRevB.102.144525} {\bibfield  {journal} {\bibinfo
   {journal} {Phys. Rev. B}\ }\textbf {\bibinfo {volume} {102}},\ \bibinfo
  {pages} {144525} (\bibinfo {year} {2020})}\BibitemShut {NoStop}%
\bibitem [{\citenamefont {Ran}\ \emph {et~al.}(2019)\citenamefont {Ran},
  \citenamefont {Eckberg}, \citenamefont {Ding}, \citenamefont {Furukawa},
  \citenamefont {Metz}, \citenamefont {Saha}, \citenamefont {Liu},
  \citenamefont {Zic}, \citenamefont {Kim}, \citenamefont {Paglione},\ and\
  \citenamefont {Butch}}]{doi:10.1126/science.aav8645}%
  \BibitemOpen
  \bibfield  {author} {\bibinfo {author} {\bibfnamefont {S.}~\bibnamefont
  {Ran}}, \bibinfo {author} {\bibfnamefont {C.}~\bibnamefont {Eckberg}},
  \bibinfo {author} {\bibfnamefont {Q.-P.}\ \bibnamefont {Ding}}, \bibinfo
  {author} {\bibfnamefont {Y.}~\bibnamefont {Furukawa}}, \bibinfo {author}
  {\bibfnamefont {T.}~\bibnamefont {Metz}}, \bibinfo {author} {\bibfnamefont
  {S.~R.}\ \bibnamefont {Saha}}, \bibinfo {author} {\bibfnamefont {I.-L.}\
  \bibnamefont {Liu}}, \bibinfo {author} {\bibfnamefont {M.}~\bibnamefont
  {Zic}}, \bibinfo {author} {\bibfnamefont {H.}~\bibnamefont {Kim}}, \bibinfo
  {author} {\bibfnamefont {J.}~\bibnamefont {Paglione}},\ and\ \bibinfo
  {author} {\bibfnamefont {N.~P.}\ \bibnamefont {Butch}},\ }\bibfield  {title}
  {\bibinfo {title} {Nearly ferromagnetic spin-triplet superconductivity},\
  }\href {https://doi.org/10.1126/science.aav8645} {\bibfield  {journal}
  {\bibinfo  {journal} {Science}\ }\textbf {\bibinfo {volume} {365}},\ \bibinfo
  {pages} {684} (\bibinfo {year} {2019})}\BibitemShut {NoStop}%
\bibitem [{\citenamefont {Jiao}\ \emph {et~al.}(2020)\citenamefont {Jiao},
  \citenamefont {Howard}, \citenamefont {Ran}, \citenamefont {Wang},
  \citenamefont {Rodriguez}, \citenamefont {Sigrist}, \citenamefont {Wang},
  \citenamefont {Butch},\ and\ \citenamefont {Madhavan}}]{2020LinJiao}%
  \BibitemOpen
  \bibfield  {author} {\bibinfo {author} {\bibfnamefont {L.}~\bibnamefont
  {Jiao}}, \bibinfo {author} {\bibfnamefont {S.}~\bibnamefont {Howard}},
  \bibinfo {author} {\bibfnamefont {S.}~\bibnamefont {Ran}}, \bibinfo {author}
  {\bibfnamefont {Z.}~\bibnamefont {Wang}}, \bibinfo {author} {\bibfnamefont
  {J.~O.}\ \bibnamefont {Rodriguez}}, \bibinfo {author} {\bibfnamefont
  {M.}~\bibnamefont {Sigrist}}, \bibinfo {author} {\bibfnamefont
  {Z.}~\bibnamefont {Wang}}, \bibinfo {author} {\bibfnamefont {N.~P.}\
  \bibnamefont {Butch}},\ and\ \bibinfo {author} {\bibfnamefont
  {V.}~\bibnamefont {Madhavan}},\ }\bibfield  {title} {\bibinfo {title}
  {{Chiral superconductivity in heavy-fermion metal UTe2 }},\ }\href
  {https://doi.org/10.1038/s41586-020-2122-2} {\bibfield  {journal} {\bibinfo
  {journal} {Nature}\ }\textbf {\bibinfo {volume} {579}},\ \bibinfo {pages}
  {523} (\bibinfo {year} {2020})}\BibitemShut {NoStop}%
\bibitem [{\citenamefont {K\"onig}\ \emph {et~al.}(2022)\citenamefont
  {K\"onig}, \citenamefont {Komijani},\ and\ \citenamefont
  {Coleman}}]{2022Konig}%
  \BibitemOpen
  \bibfield  {author} {\bibinfo {author} {\bibfnamefont {E.~J.}\ \bibnamefont
  {K\"onig}}, \bibinfo {author} {\bibfnamefont {Y.}~\bibnamefont {Komijani}},\
  and\ \bibinfo {author} {\bibfnamefont {P.}~\bibnamefont {Coleman}},\
  }\bibfield  {title} {\bibinfo {title} {Triplet resonating valence bond theory
  and transition metal chalcogenides},\ }\href
  {https://doi.org/10.1103/PhysRevB.105.075142} {\bibfield  {journal} {\bibinfo
   {journal} {Phys. Rev. B}\ }\textbf {\bibinfo {volume} {105}},\ \bibinfo
  {pages} {075142} (\bibinfo {year} {2022})}\BibitemShut {NoStop}%
\bibitem [{\citenamefont {Gor'kov}\ and\ \citenamefont
  {Rashba}(2001)}]{PhysRevLett.87.037004}%
  \BibitemOpen
  \bibfield  {author} {\bibinfo {author} {\bibfnamefont {L.~P.}\ \bibnamefont
  {Gor'kov}}\ and\ \bibinfo {author} {\bibfnamefont {E.~I.}\ \bibnamefont
  {Rashba}},\ }\bibfield  {title} {\bibinfo {title} {{Superconducting 2D system
  with lifted spin degeneracy: Mixed singlet-triplet state}},\ }\href
  {https://doi.org/10.1103/PhysRevLett.87.037004} {\bibfield  {journal}
  {\bibinfo  {journal} {Phys. Rev. Lett.}\ }\textbf {\bibinfo {volume} {87}},\
  \bibinfo {pages} {037004} (\bibinfo {year} {2001})}\BibitemShut {NoStop}%
\bibitem [{\citenamefont {Hu}\ \emph {et~al.}(2011)\citenamefont {Hu},
  \citenamefont {Jiang}, \citenamefont {Liu},\ and\ \citenamefont
  {Pu}}]{PhysRevLett.107.195304}%
  \BibitemOpen
  \bibfield  {author} {\bibinfo {author} {\bibfnamefont {H.}~\bibnamefont
  {Hu}}, \bibinfo {author} {\bibfnamefont {L.}~\bibnamefont {Jiang}}, \bibinfo
  {author} {\bibfnamefont {X.-J.}\ \bibnamefont {Liu}},\ and\ \bibinfo {author}
  {\bibfnamefont {H.}~\bibnamefont {Pu}},\ }\bibfield  {title} {\bibinfo
  {title} {{Probing anisotropic superfluidity in atomic Fermi gases with Rashba
  spin-orbit coupling}},\ }\href
  {https://doi.org/10.1103/PhysRevLett.107.195304} {\bibfield  {journal}
  {\bibinfo  {journal} {Phys. Rev. Lett.}\ }\textbf {\bibinfo {volume} {107}},\
  \bibinfo {pages} {195304} (\bibinfo {year} {2011})}\BibitemShut {NoStop}%
\bibitem [{\citenamefont {Vyasanakere}\ \emph {et~al.}(2011)\citenamefont
  {Vyasanakere}, \citenamefont {Zhang},\ and\ \citenamefont
  {Shenoy}}]{PhysRevB.84.014512}%
  \BibitemOpen
  \bibfield  {author} {\bibinfo {author} {\bibfnamefont {J.~P.}\ \bibnamefont
  {Vyasanakere}}, \bibinfo {author} {\bibfnamefont {S.}~\bibnamefont {Zhang}},\
  and\ \bibinfo {author} {\bibfnamefont {V.~B.}\ \bibnamefont {Shenoy}},\
  }\bibfield  {title} {\bibinfo {title} {{BCS-BEC crossover induced by a
  synthetic non-Abelian gauge field}},\ }\href
  {https://doi.org/10.1103/PhysRevB.84.014512} {\bibfield  {journal} {\bibinfo
  {journal} {Phys. Rev. B}\ }\textbf {\bibinfo {volume} {84}},\ \bibinfo
  {pages} {014512} (\bibinfo {year} {2011})}\BibitemShut {NoStop}%
\bibitem [{\citenamefont {Bohr}\ \emph {et~al.}(1958)\citenamefont {Bohr},
  \citenamefont {Mottelson},\ and\ \citenamefont {Pines}}]{58Bohr}%
  \BibitemOpen
  \bibfield  {author} {\bibinfo {author} {\bibfnamefont {A.}~\bibnamefont
  {Bohr}}, \bibinfo {author} {\bibfnamefont {B.~R.}\ \bibnamefont
  {Mottelson}},\ and\ \bibinfo {author} {\bibfnamefont {D.}~\bibnamefont
  {Pines}},\ }\bibfield  {title} {\bibinfo {title} {{Possible analogy between
  the excitation spectra of nuclei and those of the superconducting metallic
  state}},\ }\href {https://doi.org/10.1103/PhysRev.110.936} {\bibfield
  {journal} {\bibinfo  {journal} {Phys. Rev.}\ }\textbf {\bibinfo {volume}
  {110}},\ \bibinfo {pages} {936} (\bibinfo {year} {1958})}\BibitemShut
  {NoStop}%
\bibitem [{\citenamefont {Bohr}\ and\ \citenamefont
  {Mottelson}(1969)}]{69Bohr}%
  \BibitemOpen
  \bibfield  {author} {\bibinfo {author} {\bibfnamefont {A.}~\bibnamefont
  {Bohr}}\ and\ \bibinfo {author} {\bibfnamefont {B.~R.}\ \bibnamefont
  {Mottelson}},\ }\href@noop {} {\emph {\bibinfo {title} {Nuclear Structure}}}\
  (\bibinfo  {publisher} {W.A. Benjamin, Inc.},\ \bibinfo {address} {New York,
  USA},\ \bibinfo {year} {1969})\BibitemShut {NoStop}%
\bibitem [{\citenamefont {Ring}\ and\ \citenamefont
  {Schuck}(1980)}]{Ring_Schuck}%
  \BibitemOpen
  \bibfield  {author} {\bibinfo {author} {\bibfnamefont {P.}~\bibnamefont
  {Ring}}\ and\ \bibinfo {author} {\bibfnamefont {P.}~\bibnamefont {Schuck}},\
  }\href {https://link.springer.com/book/9783540212065} {\emph {\bibinfo
  {title} {{The Nuclear Many-Body Problems}}}},\ Texts and Monographs in
  Physics\ (\bibinfo  {publisher} {Springer-Verlag},\ \bibinfo {address} {New
  York},\ \bibinfo {year} {1980})\BibitemShut {NoStop}%
\bibitem [{\citenamefont {Broglia}\ and\ \citenamefont
  {Zelevinsky}(2013)}]{13BZ}%
  \BibitemOpen
  \bibinfo {editor} {\bibfnamefont {R.~A.}\ \bibnamefont {Broglia}}\ and\
  \bibinfo {editor} {\bibfnamefont {V.}~\bibnamefont {Zelevinsky}},\ eds.,\
  \href {http://www.worldscientific.com/worldscibooks/10.1142/8526#t=aboutBook}
  {\emph {\bibinfo {title} {Fifty Years of Nuclear BCS: Pairing in Finite
  Systems}}}\ (\bibinfo  {publisher} {World Scientific, Singapore},\ \bibinfo
  {year} {2013})\BibitemShut {NoStop}%
\bibitem [{\citenamefont {Tamagaki}(1968)}]{1968Tamagaki}%
  \BibitemOpen
  \bibfield  {author} {\bibinfo {author} {\bibfnamefont {R.}~\bibnamefont
  {Tamagaki}},\ }\bibfield  {title} {\bibinfo {title} {{Potential models of
  nuclear forces at small distances}},\ }\href
  {https://doi.org/10.1143/PTP.39.91} {\bibfield  {journal} {\bibinfo
  {journal} {Prog. Theor. Phys.}\ }\textbf {\bibinfo {volume} {39}},\ \bibinfo
  {pages} {91} (\bibinfo {year} {1968})}\BibitemShut {NoStop}%
\bibitem [{\citenamefont {Tamagaki}(1970)}]{1970Tamagaki}%
  \BibitemOpen
  \bibfield  {author} {\bibinfo {author} {\bibfnamefont {R.}~\bibnamefont
  {Tamagaki}},\ }\bibfield  {title} {\bibinfo {title} {{Superfluid State in
  Neutron Star Matter. I: Generalized Bogoliubov transformation and existence
  of $^3P_2$ gap at high density}},\ }\href
  {https://doi.org/10.1143/PTP.44.905} {\bibfield  {journal} {\bibinfo
  {journal} {Prog. Theor. Phys.}\ }\textbf {\bibinfo {volume} {44}},\ \bibinfo
  {pages} {905} (\bibinfo {year} {1970})}\BibitemShut {NoStop}%
\bibitem [{\citenamefont {Hoffberg}\ \emph {et~al.}(1970)\citenamefont
  {Hoffberg}, \citenamefont {Glassgold}, \citenamefont {Richardson},\ and\
  \citenamefont {Ruderman}}]{PhysRevLett.24.775}%
  \BibitemOpen
  \bibfield  {author} {\bibinfo {author} {\bibfnamefont {M.}~\bibnamefont
  {Hoffberg}}, \bibinfo {author} {\bibfnamefont {A.~E.}\ \bibnamefont
  {Glassgold}}, \bibinfo {author} {\bibfnamefont {R.~W.}\ \bibnamefont
  {Richardson}},\ and\ \bibinfo {author} {\bibfnamefont {M.}~\bibnamefont
  {Ruderman}},\ }\bibfield  {title} {\bibinfo {title} {{Anisotropic
  superfluidity in neutron star matter}},\ }\href
  {https://doi.org/10.1103/PhysRevLett.24.775} {\bibfield  {journal} {\bibinfo
  {journal} {Phys. Rev. Lett.}\ }\textbf {\bibinfo {volume} {24}},\ \bibinfo
  {pages} {775} (\bibinfo {year} {1970})}\BibitemShut {NoStop}%
\bibitem [{\citenamefont {Bender}\ \emph {et~al.}(2003)\citenamefont {Bender},
  \citenamefont {Heenen},\ and\ \citenamefont {Reinhard}}]{03Bender_rev}%
  \BibitemOpen
  \bibfield  {author} {\bibinfo {author} {\bibfnamefont {M.}~\bibnamefont
  {Bender}}, \bibinfo {author} {\bibfnamefont {P.-H.}\ \bibnamefont {Heenen}},\
  and\ \bibinfo {author} {\bibfnamefont {P.-G.}\ \bibnamefont {Reinhard}},\
  }\bibfield  {title} {\bibinfo {title} {Self-consistent mean-field models for
  nuclear structure},\ }\href {https://doi.org/10.1103/RevModPhys.75.121}
  {\bibfield  {journal} {\bibinfo  {journal} {Rev. Mod. Phys.}\ }\textbf
  {\bibinfo {volume} {75}},\ \bibinfo {pages} {121} (\bibinfo {year}
  {2003})}\BibitemShut {NoStop}%
\bibitem [{\citenamefont {Gandolfi}\ \emph {et~al.}(2022)\citenamefont
  {Gandolfi}, \citenamefont {Palkanoglou}, \citenamefont {Carlson},
  \citenamefont {Gezerlis},\ and\ \citenamefont {Schmidt}}]{condmat7010019}%
  \BibitemOpen
  \bibfield  {author} {\bibinfo {author} {\bibfnamefont {S.}~\bibnamefont
  {Gandolfi}}, \bibinfo {author} {\bibfnamefont {G.}~\bibnamefont
  {Palkanoglou}}, \bibinfo {author} {\bibfnamefont {J.}~\bibnamefont
  {Carlson}}, \bibinfo {author} {\bibfnamefont {A.}~\bibnamefont {Gezerlis}},\
  and\ \bibinfo {author} {\bibfnamefont {K.~E.}\ \bibnamefont {Schmidt}},\
  }\bibfield  {title} {\bibinfo {title} {{The $^1S_0$ Pairing Gap in Neutron
  Matter}},\ }\href {https://doi.org/10.3390/condmat7010019} {\bibfield
  {journal} {\bibinfo  {journal} {Condense. Matter}\ }\textbf {\bibinfo
  {volume} {7}},\ \bibinfo {pages} {19} (\bibinfo {year} {2022})}\BibitemShut
  {NoStop}%
\bibitem [{\citenamefont {Sagawa}\ \emph {et~al.}(2016)\citenamefont {Sagawa},
  \citenamefont {Bai},\ and\ \citenamefont {Col\`o}}]{Sagawa:2015zlu}%
  \BibitemOpen
  \bibfield  {author} {\bibinfo {author} {\bibfnamefont {H.}~\bibnamefont
  {Sagawa}}, \bibinfo {author} {\bibfnamefont {C.~L.}\ \bibnamefont {Bai}},\
  and\ \bibinfo {author} {\bibfnamefont {G.}~\bibnamefont {Col\`o}},\
  }\bibfield  {title} {\bibinfo {title} {{Isovector spin-singlet ($T = 1$, $S =
  0$) and isoscalar spin-triplet ($T = 0$, $S = 1$) pairing interactions and
  spin-isospin response}},\ }\href
  {https://doi.org/10.1088/0031-8949/91/8/083011} {\bibfield  {journal}
  {\bibinfo  {journal} {Phys. Scr.}\ }\textbf {\bibinfo {volume} {91}},\
  \bibinfo {pages} {083011} (\bibinfo {year} {2016})}\BibitemShut {NoStop}%
\bibitem [{\citenamefont {Oishi}\ and\ \citenamefont {Paar}(2019)}]{2019OP}%
  \BibitemOpen
  \bibfield  {author} {\bibinfo {author} {\bibfnamefont {T.}~\bibnamefont
  {Oishi}}\ and\ \bibinfo {author} {\bibfnamefont {N.}~\bibnamefont {Paar}},\
  }\bibfield  {title} {\bibinfo {title} {Magnetic dipole excitation and its sum
  rule in nuclei with two valence nucleons},\ }\href
  {https://doi.org/10.1103/PhysRevC.100.024308} {\bibfield  {journal} {\bibinfo
   {journal} {Phys. Rev. C}\ }\textbf {\bibinfo {volume} {100}},\ \bibinfo
  {pages} {024308} (\bibinfo {year} {2019})}\BibitemShut {NoStop}%
\bibitem [{\citenamefont {Oishi}\ \emph {et~al.}(2021)\citenamefont {Oishi},
  \citenamefont {Kru\v{z}i\'{c}},\ and\ \citenamefont {Paar}}]{2021Oishi_M1}%
  \BibitemOpen
  \bibfield  {author} {\bibinfo {author} {\bibfnamefont {T.}~\bibnamefont
  {Oishi}}, \bibinfo {author} {\bibfnamefont {G.}~\bibnamefont
  {Kru\v{z}i\'{c}}},\ and\ \bibinfo {author} {\bibfnamefont {N.}~\bibnamefont
  {Paar}},\ }\bibfield  {title} {\bibinfo {title} {Discerning nuclear pairing
  properties from magnetic dipole excitation},\ }\href
  {https://doi.org/10.1140/epja/s10050-021-00488-7} {\bibfield  {journal}
  {\bibinfo  {journal} {Eur. Phys. J. A}\ }\textbf {\bibinfo {volume} {57}},\
  \bibinfo {pages} {180} (\bibinfo {year} {2021})}\BibitemShut {NoStop}%
\bibitem [{\citenamefont {Vautherin}\ and\ \citenamefont
  {Brink}(1972)}]{72Vautherin}%
  \BibitemOpen
  \bibfield  {author} {\bibinfo {author} {\bibfnamefont {D.}~\bibnamefont
  {Vautherin}}\ and\ \bibinfo {author} {\bibfnamefont {D.~M.}\ \bibnamefont
  {Brink}},\ }\bibfield  {title} {\bibinfo {title} {{Hartree-Fock calculations
  with Skyrme's interaction. I. Spherical nuclei}},\ }\href
  {https://doi.org/10.1103/PhysRevC.5.626} {\bibfield  {journal} {\bibinfo
  {journal} {Phys. Rev. C}\ }\textbf {\bibinfo {volume} {5}},\ \bibinfo {pages}
  {626} (\bibinfo {year} {1972})}\BibitemShut {NoStop}%
\bibitem [{\citenamefont {Perli\ifmmode~\acute{n}\else \'{n}\fi{}ska}\ \emph
  {et~al.}(2004)\citenamefont {Perli\ifmmode~\acute{n}\else \'{n}\fi{}ska},
  \citenamefont {Rohozi\ifmmode~\acute{n}\else \'{n}\fi{}ski}, \citenamefont
  {Dobaczewski},\ and\ \citenamefont {Nazarewicz}}]{2004Jacek}%
  \BibitemOpen
  \bibfield  {author} {\bibinfo {author} {\bibfnamefont {E.}~\bibnamefont
  {Perli\ifmmode~\acute{n}\else \'{n}\fi{}ska}}, \bibinfo {author}
  {\bibfnamefont {S.~G.}\ \bibnamefont {Rohozi\ifmmode~\acute{n}\else
  \'{n}\fi{}ski}}, \bibinfo {author} {\bibfnamefont {J.}~\bibnamefont
  {Dobaczewski}},\ and\ \bibinfo {author} {\bibfnamefont {W.}~\bibnamefont
  {Nazarewicz}},\ }\bibfield  {title} {\bibinfo {title} {{Local density
  approximation for proton-neutron pairing correlations: Formalism}},\ }\href
  {https://doi.org/10.1103/PhysRevC.69.014316} {\bibfield  {journal} {\bibinfo
  {journal} {Phys. Rev. C}\ }\textbf {\bibinfo {volume} {69}},\ \bibinfo
  {pages} {014316} (\bibinfo {year} {2004})}\BibitemShut {NoStop}%
\bibitem [{\citenamefont {Bennaceur}\ \emph {et~al.}(2017)\citenamefont
  {Bennaceur}, \citenamefont {Idini}, \citenamefont {Dobaczewski},
  \citenamefont {Dobaczewski}, \citenamefont {Kortelainen},\ and\ \citenamefont
  {Raimondi}}]{Bennaceur_2017}%
  \BibitemOpen
  \bibfield  {author} {\bibinfo {author} {\bibfnamefont {K.}~\bibnamefont
  {Bennaceur}}, \bibinfo {author} {\bibfnamefont {A.}~\bibnamefont {Idini}},
  \bibinfo {author} {\bibfnamefont {J.}~\bibnamefont {Dobaczewski}}, \bibinfo
  {author} {\bibfnamefont {P.}~\bibnamefont {Dobaczewski}}, \bibinfo {author}
  {\bibfnamefont {M.}~\bibnamefont {Kortelainen}},\ and\ \bibinfo {author}
  {\bibfnamefont {F.}~\bibnamefont {Raimondi}},\ }\bibfield  {title} {\bibinfo
  {title} {Nonlocal energy density functionals for pairing and
  beyond-mean-field calculations},\ }\href
  {https://doi.org/10.1088/1361-6471/aa5fd7} {\bibfield  {journal} {\bibinfo
  {journal} {J. Phys. G: Nucl. Part. Phys.}\ }\textbf {\bibinfo {volume}
  {44}},\ \bibinfo {pages} {045106} (\bibinfo {year} {2017})}\BibitemShut
  {NoStop}%
\bibitem [{\citenamefont {Dobaczewski}\ \emph {et~al.}(1984)\citenamefont
  {Dobaczewski}, \citenamefont {Flocard},\ and\ \citenamefont
  {Treiner}}]{1984Jacek}%
  \BibitemOpen
  \bibfield  {author} {\bibinfo {author} {\bibfnamefont {J.}~\bibnamefont
  {Dobaczewski}}, \bibinfo {author} {\bibfnamefont {H.}~\bibnamefont
  {Flocard}},\ and\ \bibinfo {author} {\bibfnamefont {J.}~\bibnamefont
  {Treiner}},\ }\bibfield  {title} {\bibinfo {title}
  {{Hartree--Fock--Bogolyubov description of nuclei near the neutron-drip
  line}},\ }\href
  {https://doi.org/http://dx.doi.org/10.1016/0375-9474(84)90433-0} {\bibfield
  {journal} {\bibinfo  {journal} {Nucl. Phys. A}\ }\textbf {\bibinfo {volume}
  {422}},\ \bibinfo {pages} {103 } (\bibinfo {year} {1984})}\BibitemShut
  {NoStop}%
\bibitem [{\citenamefont {Richardson}(1972)}]{PhysRevD.5.1883}%
  \BibitemOpen
  \bibfield  {author} {\bibinfo {author} {\bibfnamefont {R.~W.}\ \bibnamefont
  {Richardson}},\ }\bibfield  {title} {\bibinfo {title} {{Ginzburg-Landau
  theory of anisotropic superfluid neutron-star matter}},\ }\href
  {https://doi.org/10.1103/PhysRevD.5.1883} {\bibfield  {journal} {\bibinfo
  {journal} {Phys. Rev. D}\ }\textbf {\bibinfo {volume} {5}},\ \bibinfo {pages}
  {1883} (\bibinfo {year} {1972})}\BibitemShut {NoStop}%
\bibitem [{\citenamefont {Masaki}\ \emph {et~al.}(2020)\citenamefont {Masaki},
  \citenamefont {Mizushima},\ and\ \citenamefont
  {Nitta}}]{PhysRevResearch.2.013193}%
  \BibitemOpen
  \bibfield  {author} {\bibinfo {author} {\bibfnamefont {Y.}~\bibnamefont
  {Masaki}}, \bibinfo {author} {\bibfnamefont {T.}~\bibnamefont {Mizushima}},\
  and\ \bibinfo {author} {\bibfnamefont {M.}~\bibnamefont {Nitta}},\ }\bibfield
   {title} {\bibinfo {title} {{Microscopic description of axisymmetric vortices
  in $^{3}P_{2}$ superfluids}},\ }\href
  {https://doi.org/10.1103/PhysRevResearch.2.013193} {\bibfield  {journal}
  {\bibinfo  {journal} {Phys. Rev. Res.}\ }\textbf {\bibinfo {volume} {2}},\
  \bibinfo {pages} {013193} (\bibinfo {year} {2020})}\BibitemShut {NoStop}%
\bibitem [{\citenamefont {Bender}\ \emph {et~al.}(2009)\citenamefont {Bender},
  \citenamefont {Bennaceur}, \citenamefont {Duguet}, \citenamefont {Heenen},
  \citenamefont {Lesinski},\ and\ \citenamefont {Meyer}}]{PhysRevC.80.064302}%
  \BibitemOpen
  \bibfield  {author} {\bibinfo {author} {\bibfnamefont {M.}~\bibnamefont
  {Bender}}, \bibinfo {author} {\bibfnamefont {K.}~\bibnamefont {Bennaceur}},
  \bibinfo {author} {\bibfnamefont {T.}~\bibnamefont {Duguet}}, \bibinfo
  {author} {\bibfnamefont {P.-H.}\ \bibnamefont {Heenen}}, \bibinfo {author}
  {\bibfnamefont {T.}~\bibnamefont {Lesinski}},\ and\ \bibinfo {author}
  {\bibfnamefont {J.}~\bibnamefont {Meyer}},\ }\bibfield  {title} {\bibinfo
  {title} {{Tensor part of the Skyrme energy density functional. II.
  Deformation properties of magic and semi-magic nuclei}},\ }\href
  {https://doi.org/10.1103/PhysRevC.80.064302} {\bibfield  {journal} {\bibinfo
  {journal} {Phys. Rev. C}\ }\textbf {\bibinfo {volume} {80}},\ \bibinfo
  {pages} {064302} (\bibinfo {year} {2009})}\BibitemShut {NoStop}%
\bibitem [{\citenamefont {Bender}\ \emph {et~al.}(2000)\citenamefont {Bender},
  \citenamefont {Rutz}, \citenamefont {Reinhard},\ and\ \citenamefont
  {Maruhn}}]{Bender2000}%
  \BibitemOpen
  \bibfield  {author} {\bibinfo {author} {\bibfnamefont {M.}~\bibnamefont
  {Bender}}, \bibinfo {author} {\bibfnamefont {K.}~\bibnamefont {Rutz}},
  \bibinfo {author} {\bibfnamefont {P.-G.}\ \bibnamefont {Reinhard}},\ and\
  \bibinfo {author} {\bibfnamefont {J.~A.}\ \bibnamefont {Maruhn}},\ }\bibfield
   {title} {\bibinfo {title} {Pairing gaps from nuclear mean-field models},\
  }\href {https://doi.org/10.1007/s10050-000-4504-z} {\bibfield  {journal}
  {\bibinfo  {journal} {Eur. Phys. J. A}\ }\textbf {\bibinfo {volume} {8}},\
  \bibinfo {pages} {59} (\bibinfo {year} {2000})}\BibitemShut {NoStop}%
\bibitem [{\citenamefont {Hinohara}(2018)}]{0954-3899-45-2-024004}%
  \BibitemOpen
  \bibfield  {author} {\bibinfo {author} {\bibfnamefont {N.}~\bibnamefont
  {Hinohara}},\ }\bibfield  {title} {\bibinfo {title} {Extending pairing energy
  density functional using pairing rotational moments of inertia},\ }\href
  {http://stacks.iop.org/0954-3899/45/i=2/a=024004} {\bibfield  {journal}
  {\bibinfo  {journal} {J. Phys. G: Nucl. Part. Phys.}\ }\textbf {\bibinfo
  {volume} {45}},\ \bibinfo {pages} {024004} (\bibinfo {year}
  {2018})}\BibitemShut {NoStop}%
\bibitem [{\citenamefont {Rohozi\ifmmode~\acute{n}\else \'{n}\fi{}ski}\ \emph
  {et~al.}(2010)\citenamefont {Rohozi\ifmmode~\acute{n}\else \'{n}\fi{}ski},
  \citenamefont {Dobaczewski},\ and\ \citenamefont
  {Nazarewicz}}]{PhysRevC.81.014313}%
  \BibitemOpen
  \bibfield  {author} {\bibinfo {author} {\bibfnamefont {S.~G.}\ \bibnamefont
  {Rohozi\ifmmode~\acute{n}\else \'{n}\fi{}ski}}, \bibinfo {author}
  {\bibfnamefont {J.}~\bibnamefont {Dobaczewski}},\ and\ \bibinfo {author}
  {\bibfnamefont {W.}~\bibnamefont {Nazarewicz}},\ }\bibfield  {title}
  {\bibinfo {title} {{Self-consistent symmetries in the proton-neutron
  Hartree-Fock-Bogoliubov approach}},\ }\href
  {https://doi.org/10.1103/PhysRevC.81.014313} {\bibfield  {journal} {\bibinfo
  {journal} {Phys. Rev. C}\ }\textbf {\bibinfo {volume} {81}},\ \bibinfo
  {pages} {014313} (\bibinfo {year} {2010})}\BibitemShut {NoStop}%
\bibitem [{\citenamefont {Bennaceur}\ and\ \citenamefont
  {Dobaczewski}(2005)}]{HFBRAD}%
  \BibitemOpen
  \bibfield  {author} {\bibinfo {author} {\bibfnamefont {K.}~\bibnamefont
  {Bennaceur}}\ and\ \bibinfo {author} {\bibfnamefont {J.}~\bibnamefont
  {Dobaczewski}},\ }\bibfield  {title} {\bibinfo {title} {{Coordinate-space
  solution of the Skyrme-Hartree--Fock--Bogolyubov equations within spherical
  symmetry. The program HFBRAD (v1.00)}},\ }\href
  {https://doi.org/https://doi.org/10.1016/j.cpc.2005.02.002} {\bibfield
  {journal} {\bibinfo  {journal} {Comp. Phys. Commun.}\ }\textbf {\bibinfo
  {volume} {168}},\ \bibinfo {pages} {96} (\bibinfo {year} {2005})}\BibitemShut
  {NoStop}%
\bibitem [{\citenamefont {Kanada-En'yo}\ and\ \citenamefont
  {Kobayashi}(2014)}]{PhysRevC.90.054332}%
  \BibitemOpen
  \bibfield  {author} {\bibinfo {author} {\bibfnamefont {Y.}~\bibnamefont
  {Kanada-En'yo}}\ and\ \bibinfo {author} {\bibfnamefont {F.}~\bibnamefont
  {Kobayashi}},\ }\bibfield  {title} {\bibinfo {title} {{Mixing of parity of a
  nucleon pair at the nuclear surface due to the spin-orbit potential in
  $^{18}\mathrm{F}$}},\ }\href {https://doi.org/10.1103/PhysRevC.90.054332}
  {\bibfield  {journal} {\bibinfo  {journal} {Phys. Rev. C}\ }\textbf {\bibinfo
  {volume} {90}},\ \bibinfo {pages} {054332} (\bibinfo {year}
  {2014})}\BibitemShut {NoStop}%
\bibitem [{\citenamefont {Yamaguchi}\ and\ \citenamefont
  {Ohashi}(2015)}]{PhysRevA.92.013615}%
  \BibitemOpen
  \bibfield  {author} {\bibinfo {author} {\bibfnamefont {T.}~\bibnamefont
  {Yamaguchi}}\ and\ \bibinfo {author} {\bibfnamefont {Y.}~\bibnamefont
  {Ohashi}},\ }\bibfield  {title} {\bibinfo {title} {{Proposed method to
  realize the $p$-wave superfluid state using an $s$-wave superfluid Fermi gas
  with a synthetic spin-orbit interaction}},\ }\href
  {https://doi.org/10.1103/PhysRevA.92.013615} {\bibfield  {journal} {\bibinfo
  {journal} {Phys. Rev. A}\ }\textbf {\bibinfo {volume} {92}},\ \bibinfo
  {pages} {013615} (\bibinfo {year} {2015})}\BibitemShut {NoStop}%
\bibitem [{\citenamefont {Beiner}\ \emph {et~al.}(1975)\citenamefont {Beiner},
  \citenamefont {Flocard}, \citenamefont {Giai},\ and\ \citenamefont
  {Quentin}}]{Beiner197529}%
  \BibitemOpen
  \bibfield  {author} {\bibinfo {author} {\bibfnamefont {M.}~\bibnamefont
  {Beiner}}, \bibinfo {author} {\bibfnamefont {H.}~\bibnamefont {Flocard}},
  \bibinfo {author} {\bibfnamefont {N.~V.}\ \bibnamefont {Giai}},\ and\
  \bibinfo {author} {\bibfnamefont {P.}~\bibnamefont {Quentin}},\ }\bibfield
  {title} {\bibinfo {title} {{Nuclear ground-state properties and
  self-consistent calculations with the Skyrme interaction: (I). Spherical
  description}},\ }\href {https://doi.org/10.1016/0375-9474(75)90338-3}
  {\bibfield  {journal} {\bibinfo  {journal} {Nucl. Phys. A}\ }\textbf
  {\bibinfo {volume} {238}},\ \bibinfo {pages} {29 } (\bibinfo {year}
  {1975})}\BibitemShut {NoStop}%
\bibitem [{\citenamefont {Chabanat}\ \emph {et~al.}(1998)\citenamefont
  {Chabanat}, \citenamefont {Bonche}, \citenamefont {Haensel}, \citenamefont
  {Meyer},\ and\ \citenamefont {Schaeffer}}]{Chabanat1998231}%
  \BibitemOpen
  \bibfield  {author} {\bibinfo {author} {\bibfnamefont {E.}~\bibnamefont
  {Chabanat}}, \bibinfo {author} {\bibfnamefont {P.}~\bibnamefont {Bonche}},
  \bibinfo {author} {\bibfnamefont {P.}~\bibnamefont {Haensel}}, \bibinfo
  {author} {\bibfnamefont {J.}~\bibnamefont {Meyer}},\ and\ \bibinfo {author}
  {\bibfnamefont {R.}~\bibnamefont {Schaeffer}},\ }\bibfield  {title} {\bibinfo
  {title} {A {Skyrme} parametrization from subnuclear to neutron star densities
  {Part} {II}. {Nuclei} far from stabilities},\ }\href
  {https://doi.org/10.1016/S0375-9474(98)00180-8} {\bibfield  {journal}
  {\bibinfo  {journal} {Nucl. Phys. A}\ }\textbf {\bibinfo {volume} {635}},\
  \bibinfo {pages} {231 } (\bibinfo {year} {1998})}\BibitemShut {NoStop}%
\bibitem [{\citenamefont {Col{\`o}}\ \emph {et~al.}(2007)\citenamefont
  {Col{\`o}}, \citenamefont {Sagawa}, \citenamefont {Fracasso},\ and\
  \citenamefont {Bortignon}}]{COLO2007227}%
  \BibitemOpen
  \bibfield  {author} {\bibinfo {author} {\bibfnamefont {G.}~\bibnamefont
  {Col{\`o}}}, \bibinfo {author} {\bibfnamefont {H.}~\bibnamefont {Sagawa}},
  \bibinfo {author} {\bibfnamefont {S.}~\bibnamefont {Fracasso}},\ and\
  \bibinfo {author} {\bibfnamefont {P.~F.}\ \bibnamefont {Bortignon}},\
  }\bibfield  {title} {\bibinfo {title} {{Spin--orbit splitting and the tensor
  component of the Skyrme interaction}},\ }\href
  {https://doi.org/http://dx.doi.org/10.1016/j.physletb.2007.01.033} {\bibfield
   {journal} {\bibinfo  {journal} {Phys. Lett. B}\ }\textbf {\bibinfo {volume}
  {646}},\ \bibinfo {pages} {227 } (\bibinfo {year} {2007})}\BibitemShut
  {NoStop}%
\bibitem [{\citenamefont {Lesinski}\ \emph {et~al.}(2007)\citenamefont
  {Lesinski}, \citenamefont {Bender}, \citenamefont {Bennaceur}, \citenamefont
  {Duguet},\ and\ \citenamefont {Meyer}}]{PhysRevC.76.014312}%
  \BibitemOpen
  \bibfield  {author} {\bibinfo {author} {\bibfnamefont {T.}~\bibnamefont
  {Lesinski}}, \bibinfo {author} {\bibfnamefont {M.}~\bibnamefont {Bender}},
  \bibinfo {author} {\bibfnamefont {K.}~\bibnamefont {Bennaceur}}, \bibinfo
  {author} {\bibfnamefont {T.}~\bibnamefont {Duguet}},\ and\ \bibinfo {author}
  {\bibfnamefont {J.}~\bibnamefont {Meyer}},\ }\bibfield  {title} {\bibinfo
  {title} {{Tensor part of the Skyrme energy density functional: Spherical
  nuclei}},\ }\href {https://doi.org/10.1103/PhysRevC.76.014312} {\bibfield
  {journal} {\bibinfo  {journal} {Phys. Rev. C}\ }\textbf {\bibinfo {volume}
  {76}},\ \bibinfo {pages} {014312} (\bibinfo {year} {2007})}\BibitemShut
  {NoStop}%
\bibitem [{\citenamefont {Hinohara}\ and\ \citenamefont
  {Nazarewicz}(2016)}]{PhysRevLett.116.152502}%
  \BibitemOpen
  \bibfield  {author} {\bibinfo {author} {\bibfnamefont {N.}~\bibnamefont
  {Hinohara}}\ and\ \bibinfo {author} {\bibfnamefont {W.}~\bibnamefont
  {Nazarewicz}},\ }\bibfield  {title} {\bibinfo {title} {{Pairing
  Nambu-Goldstone modes within nuclear density functional theory}},\ }\href
  {https://doi.org/10.1103/PhysRevLett.116.152502} {\bibfield  {journal}
  {\bibinfo  {journal} {Phys. Rev. Lett.}\ }\textbf {\bibinfo {volume} {116}},\
  \bibinfo {pages} {152502} (\bibinfo {year} {2016})}\BibitemShut {NoStop}%
\end{thebibliography}%

\end{document}